\documentclass{mn2e}

\usepackage[small]{caption}
\usepackage{graphicx}
\usepackage{multirow}
\usepackage{amssymb}
\usepackage{float}
\usepackage{lscape}

\title[Black hole spin in deep observations]{Assessing black hole spin in deep {\sl Suzaku} observations of Seyfert 1 AGN}

\author[A. Patrick et al.]{A.R. Patrick$^{1}$, J.N. Reeves$^{1}$, A.P. Lobban$^{1}$, D. Porquet$^{2}$, A.G. Markowitz$^{3}$
\\
$^{1}$Astrophysics Group, School of Physical Sciences, Keele University, Keele, Staffordshire, ST5 5BG, UK\\
$^{2}$Observatoire astronomique de Strasbourg, Universit�e de Strasbourg, CNRS, UMR 7550, 11 rue de l'Universite, F-67000 Strasbourg, France\\
$^{3}$Center for Astrophysics and Space Sciences, University of California, San Diego, M.C. 0424, La Jolla, CA, 92093-0424, USA\\
}

\pagerange{\pageref{firstpage}--\pageref{lastpage}} \pubyear{2010}

\begin{document}

\maketitle
\begin{abstract}
We present a broad-band analysis of deep {\sl Suzaku} observations of nearby Seyfert 1 AGN: Fairall 9, MCG--6-30-15, NGC 3516, NGC 3783 and NGC 4051. The use of deep observations (exposures $>200~$\,ks) with high S/N allows the complex spectra of these objects to be examined in full, taking into account features such as the soft excess, reflection continuum and complex absorption components. After a self-consistent modelling of the broad-band data (0.6-100.0\,keV, also making use of BAT data from {\sl Swift}), the subtle curvature which may be introduced as a consequence of warm absorbers has a measured affect upon the spectrum at energies $>3$\,keV and the Fe\,K region. Forming a model (including absorption) of these AGN allows the true extent to which broadened diskline emission is present to be examined and as a result the measurement of accretion disc and black hole parameters which are consistent over the full 0.6-100.0\,keV energy range.

Fitting relativistic line emission models appear to rule out the presence of maximally spinning black holes in all objects at the 90\% confidence level, in particular MCG--6-30-15 at $>99.5\%$ confidence. Relativistic Fe\,K line emission is only marginally required in NGC 3516 and not required in NGC 4051, over the full energy bandpass. Nonetheless, statistically significant broadened 6.4\,keV Fe\,K$\alpha$ emission is detected in Fairall 9, MCG--6-30-15 and NGC 3783 yielding black hole spin estimates of $a=0.67^{+0.10}_{-0.11}$, $a=0.49^{+0.20}_{-0.12}$ and $a<-0.04$ respectively, when fitted with disc emission models.
\end{abstract}


\section{Introduction}
Gaining information regarding the spin of supermassive black holes (SMBHs) in active galactic nuclei (AGN) is essential in order to understand some of the fundamental mechanisms for powering radio jets, galaxy evolution and AGN merger histories. For example, it is thought that the spin of the SMBH may be related to powering relativistic jets and it is therefore important to begin to assess the differences between radio-loud and radio-quiet AGN (Wilson \& Colbert 1995; Moderski et al. 1998; Mangalam et al. 2009) and the intrinsic properties of the SMBH and its accretion disc. Similarly, the distribution of SMBH spin or the 'average' spin could be an essential tool for distinguishing between various galaxy evolution models (Hughes \& Blandford 2003; Volonteri et al. 2005; King \& Pringle 2007; Rezzolla et al. 2008). Prolonged accretion could result in a large fraction of BHs having a high degree of spin, however, if the majority of AGN have more intermediate or low BH spin then spin-alignment during mergers could be the mechanism responsible for spin-ups or the Blandford-Znajek effect (the magnetic extraction of BH rotational energy; Blandford \& Znajek 1977) could cause the reduction of BH spin in many AGN (Berti \& Volonteri 2008).

The properties of the SMBH and the surrounding accretion disc can be measured through the analysis of the X-ray spectrum of AGN since line emission occurring from close to the black hole can become relativistically broadened (Fabian et al. 1989; Laor 1991). The shape of such a line profile allows properties such as the emissivity and the inclination of the disc to be determined, in addition to the typical inner radius of emission and the spin of the central black hole (Dov\v{c}iak et al. 2004; Brenneman \& Reynolds 2006; Dauser et al. 2010). These signatures of emission from close to the black hole are most prominent in the neutral Fe\,K$\alpha$ line at 6.4\,keV due to the high abundance and fluorescent yield of iron. In order to make the required measurements (particularly BH spin), a robust broad-band spectral model must be formed in order to assess the contribution (if any) of relativistically broadened line emission.

Recent publications have gone some way towards making tentative spin measurements in AGN through the analysis of Seyferts lacking significant amounts of intrinsic absorption (Miniutti et al. 2009; Schmoll et al. 2009; Patrick et al. 2011), or the analysis of deep observations of more complex AGN (Brenneman \& Reynolds 2006; Brenneman et al. 2011; Gallo et al. 2011). As noted in Patrick et al. (2011), the interpretation as to the origin of the soft excess can influence the obtained accretion disc parameters and BH spin. Employing a relativistically smeared reflection spectrum from the inner regions of the accretion disc (e.g. an ionized disc reflection spectrum convolved with a Kerr metric; Ross \& Fabian 2005; Brenneman \& Reynolds 2006) to model both the soft excess and broadening in the Fe\,K region yields results in which the measured parameters differ significantly when they are obtained from either the soft excess or Fe\,K region. This is particularly notable in AGN with a strong soft excess, in which case the soft excess is the driving force behind the fit, often leading to typically high spin and emissivity values i.e. $a>0.90$ and $q>4.5$. As the soft excess is often smooth or featureless, a very high degree of blurring is required. 

Alternatively employing a Comptonization origin of the soft excess (Titarchuk 1994) and seperately measuring these parameters purely from the observed line profile in the Fe\,K region yields a more moderate emissivity index and a range of spin values akin to the results found in objects without an observed soft excess i.e. $0<a<0.998$ and $q\sim2.3$ (Patrick et al. 2011). It is therefore essential that high quality data sets and deep exposures are used in order to distinguish between these scenarios and measure the true accretion disc parameters independently.

\begin{table*}
\caption{The {\sl Suzaku} Seyfert sample}
\begin{tabular}{l c c c c}
\hline
Object  & RA (J2000) & Dec (J2000) & Redshift & $N_{H}$ (Gal) ($10^{22}$cm$^{-2}$) \\ 
\hline
Fairall 9 & 01 23 45.8 & --58 48 21 & 0.0470 & 0.0316 \\
MCG--6-30-15 & 13 35 53.8 & --34 17 44 & 0.0077 & 0.0392 \\
NGC 3516 & 11 06 47.5 & +72 34 07 & 0.0088 & 0.0345 \\
NGC 3783 & 11 39 01.7 & -37 44 19 & 0.0097 & 0.0991 \\
NGC 4051 & 12 03 09.6 & +44 31 53 & 0.0023 & 0.0115 \\
\hline
\end{tabular}
\label{tab:sample}
\end{table*} 

\begin{table*}
\caption{Summary of observations for the  objects in the sample. $^1$ The observed {2--10\,keV} flux for XIS, 15--50\,keV flux for HXD and 20-100\,keV flux for BAT, in units 10$^{-11}$erg\,cm$^{-2}$\,s$^{-1}$ from the baseline model. The XIS count rates listed are per XIS. We use BAT data from the 58 month BAT catalogue (Baumgartner et al. 2010).}
\begin{tabular}{l l l c c c c c}
\hline
Object & Mission & Instrument & Date & Exposure (s) & Count rate & Flux$^{1}$ & Obs. ID \\
\hline
\multirow{3}{*}{Fairall 9} & \multirow{2}{*}{\sl Suzaku} & XIS & \multirow{2}{*}{2010/08/26} & 229296 & $1.780\pm0.002$ & 2.45 & \multirow{2}{*}{705063010} \\
& & HXD & & 162200 & $0.039\pm0.002$ & 3.51 & \\
& {\sl Swift} & BAT & -- & -- & $(4.7\pm0.2)\times10^{-4}$ & 3.34 & \\
\hline
\multirow{7}{*}{MCG--6-30-15} & \multirow{2}{*}{\sl Suzaku} & XIS & \multirow{2}{*}{2006/01/09} & 143196 & $2.862\pm0.003$ & 4.38 & \multirow{2}{*}{700007010} \\
& & HXD & & 118900 & $0.094\pm0.002$ & 4.70 & \\
& \multirow{2}{*}{\sl Suzaku} & XIS & \multirow{2}{*}{2006/01/23} & 98483 & $2.461\pm0.003$ & 3.81 & \multirow{2}{*}{700007020} \\
& & HXD & & 76800 & $0.103\pm0.003$ & 5.06 & \\
& \multirow{2}{*}{\sl Suzaku} & XIS & \multirow{2}{*}{2006/01/27} & 96691 & $2.708\pm0.003$ & 4.16 & \multirow{2}{*}{700007030} \\
& & HXD & & 83660 & $0.104\pm0.002$ & 4.98 & \\
& {\sl Swift} & BAT & -- & -- & $(6.8\pm0.3)\times10^{-4}$ & 3.86 & \\
\hline
\multirow{3}{*}{NGC 3516} & \multirow{2}{*}{\sl Suzaku} & XIS & \multirow{2}{*}{2009/10/28} & 251356 & $0.456\pm0.001$ & 1.35 & \multirow{2}{*}{704062010} \\
& & HXD & & 178200 & $0.059\pm0.001$ & 3.32 & \\
& {\sl Swift} & BAT & -- & -- & $(10.5\pm0.2)\times10^{-4}$ & 7.76 & \\
\hline
\multirow{3}{*}{NGC 3783} & \multirow{2}{*}{\sl Suzaku} & XIS & \multirow{2}{*}{2009/07/10} & 209503 & $1.942\pm0.002$ & 5.92 & \multirow{2}{*}{704063010} \\
& & HXD & & 162000 & $0.235\pm0.002$ & 12.04 & \\
& {\sl Swift} & BAT & -- & -- & $(16.6\pm0.2)\times10^{-4}$ & 11.54 & \\
\hline
\multirow{3}{*}{NGC 4051} & \multirow{2}{*}{\sl Suzaku} & XIS & \multirow{2}{*}{2008/11/06} & 274350 & $1.858\pm0.002$ & 2.45 & \multirow{2}{*}{703023010} \\
& & HXD & & 204500 & $0.062\pm0.001$ & 3.05 & \\
& {\sl Swift} & BAT & -- & -- & $(4.0\pm0.2)\times10^{-4}$ & 2.80 & \\
\hline
\end{tabular}
\label{tab:observations}
\end{table*}

This paper includes a small sample of AGN, all featuring high quality, long exposure ($>200$\,ks) observations with {\sl Suzaku} which are publically available. This is further to previous work on the analysis of a sample of relatively `bare' AGN (Patrick et al. 2011) applied to this sample of more complex AGN i.e. those featuring complex warm absorbers. {\sl Suzaku} is unique in that it offers the opportunity to gather both soft and hard X-ray data simultaneously using the XIS (Koyama et al. 2007) and HXD detectors (Takahashi et al. 2007) respectively. This enables full broad-band modelling of the 0.6-100.0\,keV spectrum, including features such as the Compton reflection hump peaking at $\sim30.0$\,keV (George \& Fabian 1991) which would otherwise be absent in spectra obtained from observations which only cover the lower energy range. Only with high energy X-ray data can the reflection component be fit simultaneous to the Fe\,K line, allowing its strength and ionization state to be determined and subsequently its contribution to the Fe\,K region. Also including time averaged BAT data (over a 58 month duration containing the {\sl Suzaku} observations) from the {\sl Swift} 58 month all sky survey (Baumgartner et al. 2010) increases the energy range up to $\sim100$\,keV. 

Alternative explanations of apparent `broad' residuals in the Fe\,K region have also been forward by some analyses in the literature, see Turner \& Miller (2009) for a review. Models consisting of complex warm absorbers and partial covering scenarios can often reduce the observed strength of any broadened residuals in the Fe\,K region (e.g. through continuum curvature imparted by bound-free transitions) and in some cases may account for these entirely. In particular, some of the objects in this sample from past observations have been modelled in such a way: MCG--6-30-15 (Miller et al. 2009), NGC 3516 (Turner et al. 2005) and NGC 3783 (Reeves et al. 2004). The complex absorption used to model these AGN is suggested to reduce the need of additional relativistic emission and indeed leads to the question whether diskline emission can be seen which is in stark constrast to typical reflection dominated models such as those used by Miniutti et al. (2007) and Brenneman \& Reynolds (2006). 

In this paper, we first fit the spectrum at soft X-ray energies and include any absorbing components which may be required by the data prior to assessing the remaining residuals in the FeK region which may be representative of diskline emission. By taking a balanced approach over a broad bandpass we aim to determine the effects of warm absorbers upon the FeK region in an attempt to discover if these broad residuals remain. Only after modelling for absorbers do we include any models which may be representative of emission from the inner regions of the accretion disc.

This paper therefore aims to assess the strength or indeed even the presence of diskline emission in the Fe\,K regions of these objects. Through modelling of absorption zones where required and the application of a variety of models and interpretations, the degeneracies involved with measuring such parameteres can be investigated and hence estimates can be placed upon BH spin and accretion disc emissivity/ inclination. We perform an analysis of these objects by forming models both with and without a partial covering component in order to assess the effect of the partial covering absorber on the iron line parameters. In later models only after modelling the absorption do we model with a dual inner (blurred) and outer (unblurred) reflector from distant material in an attempt to estimate accretion disc and BH parameters.

\section{Observations \& Data Reduction}
\subsection{Observations}
The objects included within this sample are listed in Table \ref{tab:sample} and are all the Seyfert 1, radio quiet AGN with exposures $>200$\,ks which have been observed with {\sl Suzaku} with data publically available in the {\sl Suzaku} data archive\,\footnote{http://heasarc.gsfc.nasa.gov/}. The objects in this sample are also all nearby with redshift $z<0.05$. High energy X-ray data from {\sl Swift}/BAT is also used in addition to that obtained from the HXD detector onboard {\sl Suzaku} (allowing the cross-normalization to vary), therefore the total energy range covered is 0.6-100.0\,keV. Details of the observations included are listed in Table \ref{tab:observations}. NGC 4051 has been observed on three occasions with {\sl Suzaku}, only the long November 2008 observation is included here, Lobban et al. (2011) conducted an analysis of all three simultaneously. The long 2010 {\sl Suzaku} observation of NGC 1365 is excluded from this paper due to its similarities with Seyfert 2 AGN, i.e. due to its high degree of obscuration (Maiolino et al. 2010). 

\subsection{Data reduction}
The {\sl Suzaku} data in this paper were reduced using version 6.8 of the HEASOFT data reduction and analysis package. The XIS source spectra were extracted using 3.0\arcmin circular regions centred  on the source. Background spectra were also extracted using 3.0\arcmin circular regions, this time centred on a region of the CCD not featuring any of the source or Fe\,55 calibration regions. Only data from the front-illuminated XIS\,0, 2 and 3 detectors were used due to their greater sensitivity at Fe\,K energies, however data from the back-illuminated XIS\,1 remains consistent with the front-illuminated detectors. Observations since November 2006, however, do not include data from XIS\,2 as it is no longer operational.  

MCG--6-30-15 was observed on three occasions in January 2006, this paper makes use of all three observations and the time averaged spectrum is used in the main analysis. Analysis of the individual observations yields Fe\,K line profiles which were consistent within errors and as such all three of the January 2006 observations have been coadded. The events files for each XIS\,0 observation (i.e. all three of the observations during January 2006) were loaded into \textsc{xselect} and the source and background spectra produced. This was also performed for XIS\,2 and XIS\,3 to produce single time-averaged source and background spectra for the observations of MCG--6-30-15 listed in Table \ref{tab:observations}.

The HEASOFT tool \textsc{xisrmfgen} was used to generate the XIS redistribution matrix files (rmf) and the ancilliary response files were generated using \textsc{xissimarfgen}. The data from each of the front-illuminated XIS detectors were then co-added using \textsc{mathpha}, \textsc{addrmf} and \textsc{addarf} in order to increase signal to noise. We ignore all XIS data below 0.6\,keV, above 10.0\,keV and between 1.7-2.0\,keV due to calibration uncertainties of the detectors around the Si\,K edge. 

Spectra from the HXD were extracted from the cleaned HXD/PIN events files and subsequently corrected for instrument dead time using the tool \textsc{hxddtcor}. The tuned HXD/PIN non X-ray background (NXB) events files were used for background subtraction (Fukazawa et al. 2009) and generated with $10\times$ the actual background rate in order to reduce photon noise, also with identical good time intervals (GTIs) as used in the source events. A simulated cosmic X-ray background was also produced for each observation using XSPEC v 12.6.0q with a spectral form identical to that used in Gruber et al. (1999), this was then added to the corrected NXB to form a single background file for each observation. The appropriate response and flat field files were used from the {\sl Suzaku} CALDB suitable for the respective epochs of each observation. HXD data is used over the 15.0-60.0\,keV range along with the appropriate PIN/XIS-FI cross-normalisation according to the epoch and nominal pointing position\,\footnote{http://heasarc.gsfc.nasa.gov/docs/suzaku/analysis/watchout.html}. Additional hard X-ray data from the BAT instrument onboard {\sl Swift} obtained from the 58 month BAT catalogue were also included for all objects over the range 20.0-100.0\,keV with a BAT/XIS-FI cross-normalisation allowed to vary (but typically $\sim1$).

\section{Analysis \& Results}
Spectral analysis and model fitting is performed from within XSPEC v 12.6.0q (Arnaud 1996), all models are modified by Galactic absorption which is accounted for by the \textsc{wabs} multiplicative model (Morrison \& McCammon 1983). The respective Galactic column densities were obtained using the \textsc{nh} {\sl ftool} for each source giving the weighted average $N_{\rm H}$ value of the LAB Survey of Galactic H\,{\rm I} (Kalberla et al. 2005), using abundances from Anders \& Grevesse (1989). Data is fit over the full 0.6-100.0\,keV range available, excluding those regions affected by the uncertainties in the XIS calibration mentioned above. The $\chi^{2}$ minimisation technique is used throughout, all errors are quoted at the 90\% confidence level ($\triangle\chi^{2}=2.71$ for one interesting parameter) and include statistical and not instrumental systematic errors. Where the significance of components is quoted in terms of $\triangle\chi^{2}$, the component in question has been removed from the model and the data refit to ensure that the order in which components are added to the model does not play a role in the quoted statistical significance. Throughout this paper, a positive $\triangle\chi^{2}$ corresponds to a worsening in the fit, whereas a negative $\triangle\chi^{2}$ corresponds to an improvement in the fit. 

\subsection{Spectral Models}
\subsubsection{Modelling of the soft excess}
There are a number of possible origins of the soft excess observed in AGN X-ray spectra. A simple method of modelling the soft excess is as a black body to replicate direct thermal emission from the accretion disc (Malkan \& Sargent 1982). However, the relatively constant temperature of the soft excess is in disagreement with the typical accretion disc properties for a SMBH which should scale with $M_{\rm BH}^{-1/4}$ - more typical of an intermediate mass black hole e.g. ${\rm kT}\sim100-200$\,eV (Gierli\'{n}ski \& Done 2004). A variation upon this concept is that the soft excess originates from the inverse-Compton scattering of the EUV photons from the accretion disc in a hot plasma above the disc, which would allow for a large variation in photon seed temperature, yet it would still produce a relatively constant photon temperature  such as the \textsc{compTT} model (Titarchuk 1994). Such a model has been successful in modelling the soft excess in a sample of PG quasars (see Porquet et al. 2004). 

Departing from an origin of the soft excess due to emission from the accretion disk, an alternative explanation is an atomic origin. In this scenario soft X-ray emission lines are responsible for the soft excess, however since these are discrete features and the observed soft excess is typically smooth, the reflection continuum must be relativistically blurred and hence originate from the inner regions of the accretion disc (Ross, Fabian \& Ballantyne 2002). This interpretation of the soft excess has also proved successful in some AGN such as Ark 120, Fairall 9 and MCG--6-30-15 (Ballantyne, Vaughan \& Fabian 2003; Crummy et al. 2006; Schmoll et al. 2009; Nardini et al. 2010).

An alternative origin for the soft excess as suggested by Gierli\'{n}ski \& Done (2004) is relativistically smeared absorption whereby the discrete atomic features are masked by high velocities. This would result in a 'hole' in the spectrum creating an observed soft excess and hardening at higher energies. 

\subsubsection{Compton reflection}
Following the availability of hard X-ray data from {\sl Suzaku's} HXD and {\sl Swift's} BAT detectors features such as the Compton reflection hump at $\sim30$\,keV can be measured in the spectra of these AGN. Using an ionized reflection model such as \textsc{reflionx} (Ross \& Fabian 2005) allows properties such as the ionization state, the iron abundance ($Z_{\rm Fe}$) of the disc and strength of the reflection component to be measured.

Narrow fluorescent emission lines in AGN spectra result from reflection off distant material and as such an {\sl unblurred} reflection component should be included as part of a baseline model in AGN spectra. Previous studies of the iron line regions in Seyfert AGN by Bianchi et al. (2004), Nandra et al. (2007) and Patrick et al. (2011) also suggest that ionized species of iron are relatively common at energies of 6.7\,keV and 6.97\,keV for Fe\,{\rm XXV} and Fe\,{\rm XXVI} respectively. Neutral narrow Fe\,K$\alpha$ and the accompanying Fe\,K$\beta$ emission at 6.4\,keV and 7.056\,keV respectively are ubiquitous in AGN spectra (Nandra et al. 2007).

To account for the distant reflection component in these AGN, we use the \textsc{pexrav} (Magdziarz \& Zdziarski 1995) neutral reflection model initially which is then replaced in later models with an unblurred \textsc{reflionx} ionized reflection model with the input photon index $\Gamma$ tied to that of the intrinsic powerlaw. The \textsc{reflionx} model includes soft narrow emission lines in addition to narrow Fe\,K$\alpha$. However, Fe\,K$\beta$ emission with flux $F_{\rm K\beta}=0.13\times\,F_{\rm K\alpha}$ is not included self-consistently in \textsc{reflionx} and as such is modelled using a narrow Gaussian of fixed width $\sigma=0.01$\,keV and with flux fixed at the value obtained during the initial parametrisation. Additional narrow ionized emission lines e.g. at 6.7\,keV and 6.97\,keV may also be required and are modelled using a narrow Gaussian with width fixed as above.

\subsubsection{Warm absorption}
The presence of an X-ray absorber primarily affects the X-ray spectrum of soft energies, however, with higher column densities the warm absorber may add subtle spectral curvature to the spectrum above 2.5\,keV. In particular, highly ionized absorption zones can affect the measured line parameters and strength of the observed broad red-wing in the Fe\,K region (Reeves et al. 2004; Turner at al. 2005). In previous observations of these objects (bar Fairall 9) the presence of complex absorption zones has been well documented (Kaspi et al. 2000; Yaqoob et al. 2005; McKernan et al. 2007; Markowitz et al. 2008; Miller et al. 2008; Lobban et al. 2011).

To model the soft X-ray warm absorber components in this paper, we use the \textsc{xstar} generated (Kallman et al. 2004) grid illuminated by a photon index of $\Gamma=2$, abundances fixed at solar values (except Ni, which was set to zero) and with a turbulent velocity of $200\,{\rm km}\,{\rm s}^{-1}$. This grid is well suited to accounting for typical absorption zones due to its wide range in column density ($5\times10^{18}{\rm cm}^{-2}<N_{\rm H}<5\times10^{24}{\rm cm}^{-2}$) and ionization parameter ($0<{\rm log}\,\xi<5$). During the fitting process absorption zones are added as required, in some objects more than one zone may be statistically required.

\subsubsection{Highly ionized absorption}
In the event that absorption features in the Fe\,K region are found (such as 1s-2p resonance lines from Fe\,{\rm XXV} or Fe\,{\rm XXVI}), these will be accounted for using a model representative of a highly ionized absorption zone. We use an \textsc{xstar} generated grid with a turbulent velocity of $1000\,{\rm km}\,{\rm s}^{-1}$ (to better model the observed equivalent widths of the lines), an input continuum of $\Gamma=2.0$ and range in column density and ionization parameter of $1\times10^{20}{\rm cm}^{-2}<N_{\rm H}<1\times10^{24}{\rm cm}^{-2}$ and $0<{\rm log}\,\xi<6$ respectively.

\subsubsection{Partial covering}
Some models used in the analysis of these AGN use partial covering geometries whereby a fraction of the observed X-rays are absorbed by a surrounding gas in the line-of-sight (in addition to typical fully-covering absorbers) as described in Section 3.1.3, while some fraction of the continuum `leaks' through and is unaffected by the partially covering material. The partial coverer used here takes the form of (powerlaw + WA*powerlaw) with the photon index of both the powerlaw components tied and normalisation free to vary. The parameters of the warm absorber (ionization and column density) are also allowed to vary. The column density of the partially covering medium can have a significant affect upon the spectrum. For example, high column partial coverers ($N_{\rm H}\sim10^{24}\,{\rm cm}^{-2}$) predominantly affect the hard X-ray energies and are used here to supplement the distant reflection component in cases where a hard excess remains. Lower column density partial coverers may greatly affect spectral curvature at lower X-ray energies and in some cases can entirely remove any `broad' residuals in the Fe\,K region (Miller et al. 2009).

\begin{figure*}
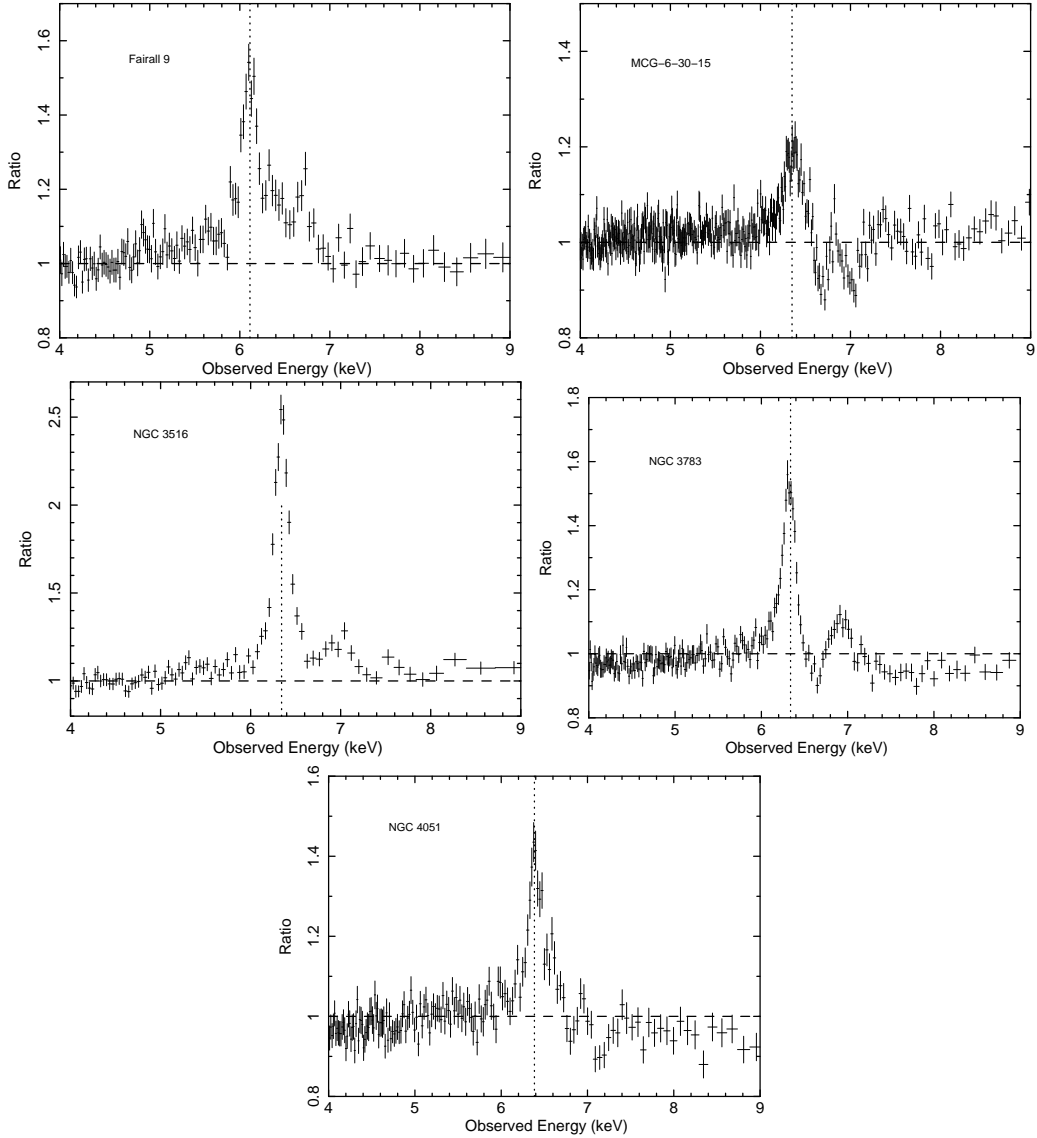

\rotatebox{-90}{\includegraphics[width=5cm]{fairall9_long_FeK_none.ps}}
\rotatebox{-90}{\includegraphics[width=5cm]{mcg6_jan06_FeK_none.ps}}
\rotatebox{-90}{\includegraphics[width=5cm]{ngc3516_long_FeK_none.ps}}
\rotatebox{-90}{\includegraphics[width=5cm]{ngc3783_long_FeK_none.ps}}
\rotatebox{-90}{\includegraphics[width=5cm]{ngc4051_obs2_FeK_none_rebin.ps}}
\caption{Ratio plots of the 4-9\,keV residuals without any modelling of the reflection component i.e. the Fe\,K region is left totally unmodelled with only warm absorption at soft X-ray energies being taken into account in addition to the intrinsic powerlaw and \textsc{compTT} (where required). The vertical dashed lines represent 6.4\,keV in the observed frames.}
\label{fig:FeK_none}
\end{figure*}

\subsection{Initial parametrization and baseline model}
An initial parametrization of the spectra is made in order to assess properties such as the strength of the reflection continuum and the strength of any emission or absorption features which may be present in the Fe\,K region. It should be noted that all objects in this sample (with the exclusion of Fairall 9) feature complex intrinsic absorption components. Once these basic components have been established, resulting in a reasonable fit to the data, the Fe\,K region can then be assessed as to the presence of any additional features. For example, narrow emission lines such as Fe\,{\rm XXV} and Fe\,{\rm XXVI}, or indeed the need for a highly ionized absorption zone (see Reeves et al. 2004 for Fe\,{\rm XXV} absorption in an {\sl XMM-Newton} observation of NGC 3783) may be required to model the Fe\,K region.

The strength of the reflection component is initially measured using the \textsc{pexrav} neutral reflection model (Magdziarz \& Zdziarski 1995) with the reflection fraction $R=\Omega/2\pi$ left free to vary (where $R=1$ denotes reflection from material subtending $2\pi$\,sr as seen from the illuminating source). The cut-off energy for the \textsc{pexrav} component is fixed at 300\,keV for consistency with the built-in cut-off in the later used \textsc{reflionx} model (similarly the powerlaw cut-off is fixed at 300\,keV) and $cos\,i=0.87$ throughout. Narrow Gaussians of fixed width (10\,eV) are added where appropriate to model neutral and ionized emission or absorption features. Fe\,K$\beta$ emission at 7.056\,keV is accounted for with fixed line energy and flux $F_{\rm K\beta}=0.13\times\,F_{\rm K\alpha}$. The flux of the Fe\,K$\beta$ component measured here is frozen and carried forward into later models since it is not included within \textsc{reflionx}. Any broad residuals in the Fe\,K region are modelled using a broad Gaussian with $\sigma$ width free to vary. The parameters obtained during this initial parametrisation are listed in Table \ref{tab:parametrisation}.  

The baseline model is intended to model the entire 0.6-100.0\,keV range in full to account for features such as any soft excess which may be present and distant reflection in a more self consistent manner than in the initial parametrization. For example, by replacing \textsc{pexrav} with \textsc{reflionx} to model reflection and using high ionization \textsc{xstar} grids to model absorption in the Fe\,K region. This baseline model does not, however, include any components which may be representative of emission occurring from regions close to the central black hole i.e. any broadened lines or observed red-wing are left unmodelled for the purposes of this first step, or null hypothesis.  

A Comptonization origin of the soft excess is assumed in this model, using \textsc{compTT} with a soft photon input temperature fixed at 0.02\,keV. This is quantitively similar to adding a steeper soft powerlaw component.  The unblurred distant reflection component is accounted for initially  by \textsc{pexrav} (plus Gaussians to model the Fe\,K region) and then as described in Section 3.1.2 above for the baseline model. Additional soft X-ray emission lines (such as O\,{\rm VII}, O\,{\rm VIII} etc) are added on an ad-hoc basis where required, again with fixed width $\sigma=0.01$\,keV. The warm absorber component is treated as stated in Section 3.1.3 with successive zones of an \textsc{xstar} grid added as statistically required. As expected from previous studies of these objects, MCG--6-30-15, NGC 3516, NGC 3783 and NGC 4051 all require at least one ionized absorption zone. In Fairall 9, however, the Galactic line-of-sight absorption is sufficient to describe the total absorption at soft X-ray energies. The warm absorber in these objects may indeed be more complex than modelled here, although the main aim is to derive and measure black hole and accretion disc properties from the Fe\,K region, given a consistent model to the broadband {\sl Suzaku} data. Once the baseline model has been parametrised we can then proceed to examine the nature of the residuals in the Fe\,K region and the strength of any broad red-wing indicative of emission from the inner regions of the accretion disc. 

\subsubsection{Fairall 9}
Fairall 9 is well modelled using an approach similar to that used in Patrick et al. (2011), consisting of \textsc{compTT} to model the soft excess plus powerlaw and an unblurred \textsc{reflionx} to account for distant reflection. The broadband spectral properties in this long observation are very similar to those in the shorter 2007 observation, however, the initial parametrisation measures a reflection fraction of $R=1.55^{+0.26}_{-0.24}$ compared to $R=0.52^{+0.20}_{-0.18}$ in the previous 2007 observation. A broad Gaussian improves the fit to the Fe\,K region with $\triangle\chi^{2}=-13$ and parameters which are consistent with Patrick et al. (2011), see Table \ref{tab:parametrisation}. Fitting both the 2010 and 2007 observations simultaneously and allowing the reflection fraction and intrinsic powerlaw to vary gives consistent results with both analyses and a smaller difference between the reflection fractions for each observation. Forcing the powerlaw $\Gamma$ to be the same in both observations provides a good fit to the XIS spectra of both observations with the majority of the residuals remaining in the fit to the HXD/PIN data. 

Due to the increased signal-to-noise in this observation, an additional narrow emission line due to O\,{\rm VIII}\,Ly$\alpha$ is observed at $0.66^{+0.01}_{-0.02}$\,keV with an improvement of $\triangle\chi^{2}=-35$ for a further two free parameters. However, since the \textsc{reflionx} model includes some soft X-ray emission lines, in this high quality data it is these lines which appear to determine the obtained reflection parameters and are therefore detrimental to the overall fit. In particular the hard X-ray data from the HXD is poorly fit. In order to resolve this issue, a neutral absorber at the redshift of the source ($z=0.047$) is included to only absorb the soft X-ray emission lines included within \textsc{reflionx}. This results in an improvement of $\triangle\chi^{2}=-109$ in the overall fit with column density $N_{\rm H}=39.59^{+8.11}_{-6.19}\times10^{22}\,{\rm cm}^{-2}$. 
Prior to including any diskline models indicative of relativistic emission, ionized narrow emission lines which may result from Fe\,{\rm XXV} and Fe\,{\rm XXVI} are found at $6.68^{+0.03}_{-0.01}$\,keV and $7.00^{+0.05}_{-0.05}$\,keV respectively (see Table \ref{tab:baseline}). These lines were also found in the shorter 2007 observation of Fairall 9 (Schmoll et al. 2009; Patrick et al. 2011) with consistent energy and $EW$.

\begin{figure*}
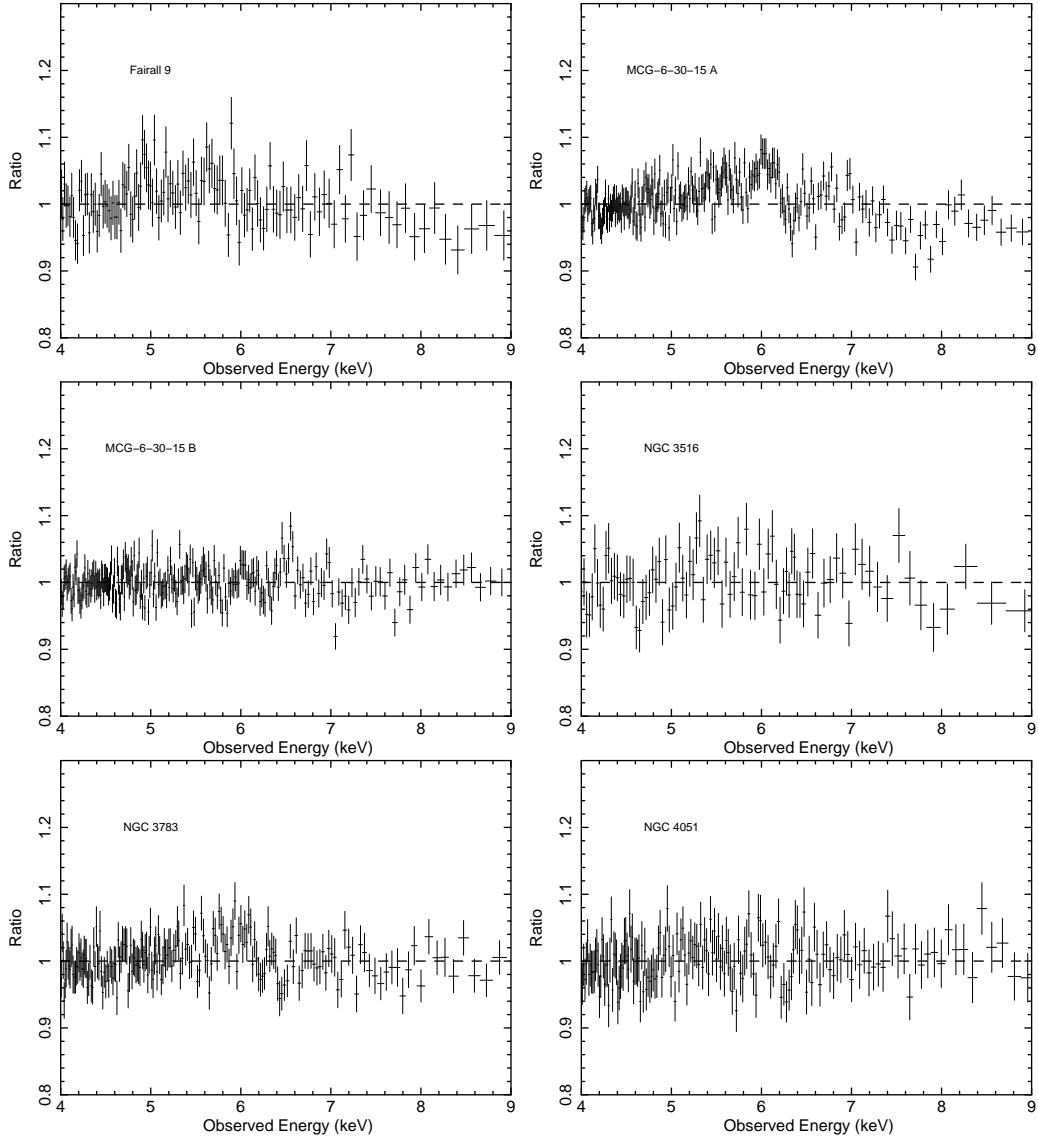

\rotatebox{-90}{\includegraphics[width=5cm]{fairall9_long_nobroad.ps}}
\rotatebox{-90}{\includegraphics[width=5cm]{mcg6_baseline_nopc.ps}}
\rotatebox{-90}{\includegraphics[width=5cm]{mcg6_jan06_nobroad.ps}}
\rotatebox{-90}{\includegraphics[width=5cm]{ngc3516_long_nobroad.ps}}
\rotatebox{-90}{\includegraphics[width=5cm]{ngc3783_long_nobroad.ps}}
\rotatebox{-90}{\includegraphics[width=5cm]{ngc4051_obs2_nobroad_rebin.ps}}
\caption{Ratio plots of the XIS spectra to the baseline model (i.e. Table \ref{tab:baseline}, without including a broad Fe\,K$\alpha$) revealing excesses at energies red-ward of 6.4\,keV in some objects. Plot MCG--6-30-15 A shows the residuals to the Fe\,K region without the use of a high column partial covering component, plot MCG--6-30-15 B shows the residuals of the complete baseline model as per the other objects. Note that the high column partial coverer significantly reduces the strength of the red-wing in MCG--6-30-15.}
\label{fig:baseline}
\end{figure*}

\subsubsection{MCG--6-30-15}
In this analysis we take the time averaged spectrum of MCG--6-30-15 during the January 2006 observations, preliminary analysis reveals that the dervied Fe\,K region parameters are consistent for all three observations and as such they are co-added as per Miniutti et al. (2007). Most apparent is the presence of a small soft and a hard excess, in addition to the requirement for some degree of warm absorption. A very poor fit to the data is found without any warm absorption with $\chi^{2}_{\nu}>60$. A single low column ($N_{\rm H}\sim7.5\times10^{21}\,{\rm cm}^{-2}$, ${\rm log}\,\xi\sim1.6$) warm absorption zone improves the fit to $\chi^{2}_{\nu}\sim3.2$. A second low column ($N_{\rm H}\sim2.4\times10^{21}\,{\rm cm}^{-2}$) ionized absorption zone improves the fit by $\triangle\chi^{2}=-1284$ to $\chi^{2}_{\nu}\sim2.5$ with both zones having ionizations of ${\rm log}\,\xi\sim0.96$ and ${\rm log}\,\xi\sim1.94$. 

In order to parametrise the reflection in MCG--6-30-15, we initially use the \textsc{pexrav} neutral reflection model. This yields a reflection fraction of $R=0.79^{+0.12}_{-0.08}$ with a good fit to the HXD and BAT data. This then allows us to measure the strength of features in the complex Fe\,K region prior to employing a more self-consistent reflection model such as \textsc{reflionx}. There is strong neutral Fe\,K$\alpha$ and what may be ionized Fe\,{\rm XXV} emission improving the fit by $\triangle\chi^{2}=-331$ and -51 respectively (see Table \ref{tab:parametrisation}); it is, however, possible the ionized feature may be the blue-wing to a broadened Fe\,K line profile. We also find ionized Fe\,{\rm XXVI} emission which is fixed at 6.97\,keV throughout the fits. In addition to this, very clear in the Fe\,K region are two ionized absorption features at $6.76^{+0.02}_{-0.02}$\,keV and $7.05^{+0.02}_{-0.02}$\,keV (see Figure \ref{fig:FeK_none}) indicative of the presence of a highly ionized absorber (as also found in a {\sl Chandra} HETG observation of MCG--6-30-15 by Young et al. 2005). Subsequent to modelling these narrow features there is still a poor fit to the data ($\chi^{2}_{\nu}\sim1.3$), however there are strong residuals remaining extending from $\sim5.0-6.4$\,keV. Including a broad Gaussian to model this feature results in a line centred at $5.93^{+0.07}_{-0.14}$\,keV and $EW=149^{+21}_{-9}$\,eV. Previous analyses of this object at energies $>3$\,keV found a much stronger broadened iron line, for example Miniutti et al. (2007) (when modelling with a `double Gaussian' to account for both the red and blue-wings) find consistent line energy and $\sigma_{\rm width}$ but a much higher cumulative $EW=320^{+45}_{-45}$\,eV in an analysis of the same time averaged data used here. 

Replacing the \textsc{pexrav} model and narrow 6.4\,keV Gaussian with \textsc{reflionx} (Fe\,K$\beta$ emission with flux fixed as in Table \ref{tab:parametrisation}) to form the baseline model provides a poor fit to the data ($\chi^{2}_{\nu}=1.82$). Despite the two absorption zones and soft emission included with \textsc{reflionx}, further soft narrow emission and absorption features are required. These can be attributed to O\,{\rm VIII}\,Ly$\alpha$, O\,{\rm VII} RRC, Ne\,{\rm IX} resonance emission and S\,{\rm XV/ XVI} 1s-2p absorption lines at $2.37^{+0.01}_{-0.01}$\,keV and $2.77^{+0.04}_{-0.08}$\,keV (see Table \ref{tab:soft} for a summary of narrow soft X-ray lines in each object). 

After using \textsc{reflionx} to model the reflection, there is still an excess at hard X-ray energies. Since the baseline model does not assume any emission from the inner regions of the accretion disc, the additional hard excess is therefore accounted for using a high column density partial covering model to supplement the \textsc{reflionx} hard X-ray component. The resulting model provides a reasonable fit to the data at both soft and hard X-ray energies with $\chi^{2}_{\nu}\sim1.26$ and an obscuring column density $N_{\rm H}\sim3.4\times10^{24}\,{\rm cm}^{-2}$ resulting in a covering fraction of $C_{\rm frac}\sim50\%$. It should be noted that due to the high column of the partial coverer, there is some effect to the residuals in the Fe\,K region and as such using such a model to account for the hard excess influences findings relating to the strength of any relativistic emission which may be present (see Figure \ref{fig:baseline} and Section 3.3.1 for more details). A good fit is obtained to the reflection continuum and neutral Fe\,K$\alpha$ line with purely Solar iron abundance.

The remaining residuals in the Fe\,K region indicate the {\sl possibility} of a {\sl moderate} red-wing below 6.4\,keV extending down to $\sim5$\,keV in addition to narrow ionized emission from Fe\,{\rm XXVI} and strong absorption due to Fe\,{\rm XXV} and Fe\,{\rm XXVI}. Following the procedure stated in Section 3.1.4, we use an \textsc{xstar} generated grid to account for the highly ionized absorption, as a consequence of this some of the observed red-wing may become weaker as noted in Reeves et al. (2004). This improves the fit further by $\triangle\chi^{2}=-298$ for an additional 3 free parameters and the residuals show no further absorption. The highly ionized absorption parameters found here are consistent with those found in past studies of MCG--6-30-15 (Miller et al. 2008; 2009). The inclusion of weak Fe\,{\rm XXVI}\,Ly$\alpha$ emission with a reduction of $\triangle\chi^{2}=-7$ finalises the baseline model.

The baseline model (Table \ref{tab:baseline}), therefore, provides a reasonable fit to the time averaged January 2006 without the need for any {\sl strong} emission from the inner regions of the accretion disc with a final fit statistic of $\chi^{2}_{\nu}=1.11$. However, some previous analyses of this object have found no such feature i.e. partial covering and absorption components alone have been claimed to give an acceptable fit. For example, Miller et al. (2009) model MCG--6-30-15 with `clumpy' absorption zones and achieve a good fit to the data without the need for the addition of a diskline component representative of emission from the inner regions of the accretion disc. In the present analysis, however, we do find that there is still some indication of a moderate red-wing which may be due to a relativistically broadened component (see Figure \ref{fig:baseline}; this is later fit with a \textsc{kerrdisk} line profile with an improvement of $\triangle\chi^{2}=-59$, see Section 3.3 and Table \ref{tab:kerrdisk}).

\subsubsection{NGC 3516}
Taking a similar approach as above, we again consider the time averaged data from the October 2009 deep observation ($\sim251$\,ks) of NGC 3516. Initially modelling the broadband spectrum with a simple Galactically absorbed powerlaw and a single warm absorption zone, a poor fit is obtained with $\chi^{2}_{\nu}\sim6.6$ and photon index of $\Gamma\sim1.7$. The most striking residuals are at hard energies and the narrow Fe\,K$\alpha$ core. There is no indication of a soft excess as found in an analysis of the 2005 observation by Markowitz et al. (2008) although NGC 3516 is highly variable. 

Parametrising the reflection continuum with \textsc{pexrav} results in a reflection fraction of $R=0.87^{+0.32}_{-0.22}$ and a relatively simple Fe\,K region with a very strong narrow Fe\,K$\alpha$ line with $EW=229^{+10}_{-52}$\,eV improving the fit by $\triangle\chi^{2}=-1396$ in addition to narrow Fe\,{\rm XXV} emission (Figure \ref{fig:FeK_none}). A broad Gaussian is added to the model which improves the fit by $\triangle\chi^{2}\sim-17$ and $EW=55^{+29}_{-28}$\,eV as noted in Table \ref{tab:parametrisation}. 

Replacing the reflection continuum with that from \textsc{reflionx} (and no longer modelling the Fe\,K region with Gaussians), however, leaves additional excesses at hard X-ray energies and requires the introduction of a high column partial covering component, consistent with the method used above for MCG--6-30-15. This makes a further significant improvement to the fit ($\triangle\chi^{2}\sim-94$) with a covering column density of $N_{\rm H}\sim1.2\times10^{23}\,{\rm cm}^{-2}$, ionization ${\rm log}\,\xi\sim0.56$ and a covering fraction of $C_{\rm frac}\sim21\%$, resulting in a reasonable fit to the broadband data. 

\begin{figure*}
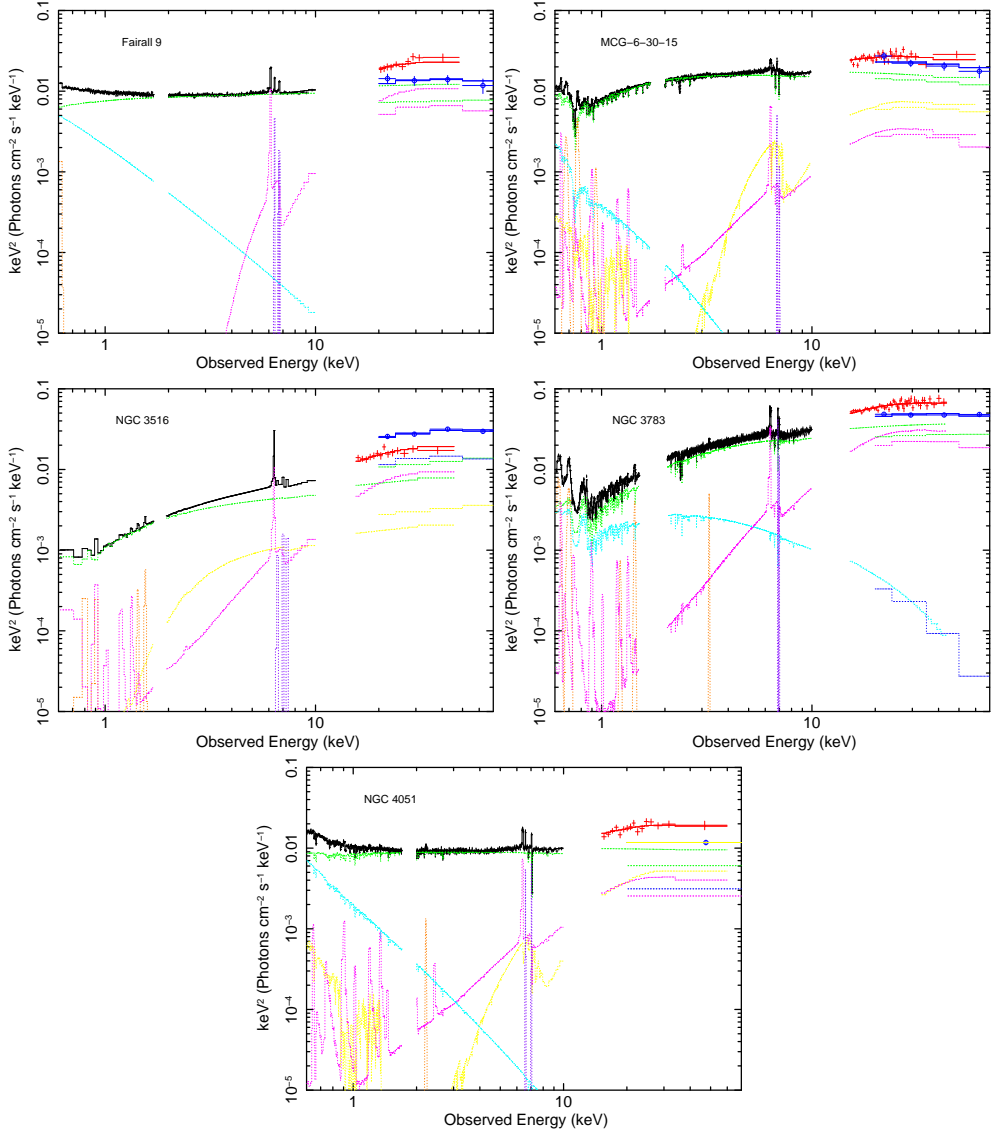

\rotatebox{-90}{\includegraphics[width=5cm]{fairall9_long_baseline_eeuf.ps}}
\rotatebox{-90}{\includegraphics[width=5cm]{mcg6_baseline_eeuf.ps}}
\rotatebox{-90}{\includegraphics[width=5cm]{ngc3516_baseline_eeuf.ps}}
\rotatebox{-90}{\includegraphics[width=5cm]{ngc3783_baseline_eeuf.ps}}
\rotatebox{-90}{\includegraphics[width=5cm]{ngc4051_baseline_eeuf_rebin.ps}}
\caption{$\nu$F$\nu$ plots of the baseline model including any absorption components: powerlaw (green), \textsc{compTT} (light blue), \textsc{reflionx} (purple), partial coverer (yellow), narrow Gaussians to model additional Fe\,K emission (dark purple) and soft emission lines (orange) where required. HXD data is in red and BAT data from {\sl Swift} is represented by blue circles. }
\label{fig:baseline_eeuf}
\end{figure*}

There are residuals at soft energies possibly originating from O\,{\rm VII} recombination, Ne\,{\rm IX}, Mg\,{\rm XII} and what may be Mg\,{\rm XI}  1s-3p emission (see Table \ref{tab:soft}), in addition to these there is a narrow emission feature observed at $7.42^{+0.05}_{-0.06}$\,keV due to Ni\,K$\alpha$ with $EW=37^{+13}_{-14}$\,eV. Analysis of any remaining residuals in the Fe\,K region do not indicate any highly ionized absorption at Fe\,{\rm XXV}/{\rm XXVI} energies. Narrow emission at $6.61^{+0.08}_{-0.07}$\,keV is observed and modelled with a narrow Gaussian, improving the fit by $\triangle\chi^{2}=-8$ for an additional two free parameters. This baseline model is a very good fit to the data with $\chi^{2}_{\nu}=1.053$ without any relativistic line profile. There is little indication of any strong red-wing below 6.4\,keV after full modelling of the broadband continuum (Figure \ref{fig:baseline} \& Figure \ref{fig:baseline_eeuf}). However, it is possible that the partial covering component may be reducing observed spectral curvature which may in fact result from relativistic emission (see Section 3.3.1). In a previous analysis a 2005 {\sl Suzaku} observation of NGC 3516, a relativistically broadened line was required (Markowitz et al. 2008), however the 2-10\,keV flux is higher by a factor $\sim1.8$ in the 2005 observation. 

\subsubsection{NGC 3783}
In NGC 3783 there is a clear soft and hard X-ray excess along with some degree of intrinsic absorption, these are initially modelled as above using the \textsc{compTT}, \textsc{pexrav} and warm absorber components in addition to the intrinsic powerlaw and Galactic absorption (using \textsc{wabs}). Two absorption zones are required, including a low column, low ionization zone ($N_{\rm H}\sim1\times10^{21}\,{\rm cm}^{-2}$, ${\rm log}\,\xi\sim0.7$, see Table \ref{tab:baseline}) consistent with Reeves et al. (2004). This still yields a poor fit to the data with $\chi^{2}_{\nu}\sim2.4$, however, there are clear residuals at soft energies due to narrow emission.

The observed soft X-ray emission lines are included using narrow Gaussians due to O\,{\rm VIII}\,Ly$\alpha$, O\,{\rm VII} recombination, Ne\,{\rm IX} recombination and Mg\,{\rm XII}\,Ly$\alpha$ (see Table \ref{tab:soft}). In addition to this there is a narrow emission feature at $3.28^{+0.03}_{-0.02}$\,keV which may be due to Ar\,{\rm XVIII}, improving the fit by $\triangle\chi^{2}=-26$ and a significant narrow absorption feature at $2.39^{+0.01}_{-0.01}$\,keV due to 1s-2p S\,{\rm XV} with an improvement of $\triangle\chi^{2}=-163$ (despite the inclusion of two warm absorption zones, however, parameters associated with the S\,{\rm XV} absorption may be influenced by Au\,M-edge calibration uncertainties). 

Following on from the Reeves et al. (2004) analysis of a long {\sl XMM-Newton} observation of NGC 3783 in 2001, we can confirm here the presence of strong Fe\,{\rm XXV} absorption in the Fe\,K region in addition to weaker absorption due to less ionized Fe at $\sim6.5$\,keV (Figure \ref{fig:FeK_none}). Modelling this absorption with two simple Gaussians results in an improvement by $\triangle\chi^{2}=-38$ and $\triangle\chi^{2}=-7$ respectively. However, to form a physical model to account for the observed complexity in the Fe\,K region we use a high ionization \textsc{xstar} grid (as in MCG--6-30-15), whilst not as good a fit as the two simple Gaussians, the fit is improved by $\triangle\chi^{2}=-18$. The high ionization zone successfully models the Fe\,{\rm XXV} absorption with the parameters given in Table \ref{tab:baseline}, although the less ionized Fe absorption at $\sim6.5$\,keV is left relatively unmodelled. Attempting to model this feature with a second highly ionized absorption zone is unsuccessful due to its relatively low statistical weight and as such the absorption feature at $\sim6.5$\,keV is left unmodelled. 

An additional narrow Fe\,{\rm XXVI} emission line is included ($\triangle\chi^{2}=-45$) to give the baseline model for NGC 3783 which provides a good fit to the data ($\chi^{2}_{\nu}=1.083$). However, the inclusion of a highly ionized absorption zone still leaves a moderate red-wing below 6.4\,keV extending down to $\sim5$\,keV (Figure \ref{fig:baseline}). This indicates the possible presence of statistically significant relativistic emission from the inner regions of the accretion disc (later fit with a \textsc{kerrdisk} line profile with an improvement of $\triangle\chi^{2}=-32$, see Section 3.3 and Table \ref{tab:kerrdisk}).
%

\subsubsection{NGC 4051}
The X-ray spectrum of NGC 4051 is very complex as well as variable, differing between all three observations to date with {\sl Suzaku}. An in depth study of these observations with the aid of HETG data from {\sl Chandra} is conducted in Lobban et al. (2011) and we take note of their findings in our modelling of the 2008 deep observation (Obs ID:703023010) during a period of relatively high flux for this source $F_{2-10\,{\rm keV}}=2.45\times10^{-11}$erg\,cm$^{-2}$\,s$^{-1}$. There are clear soft and hard excesses which are modelled with \textsc{compTT} and \textsc{pexrav} respectively, along with two low column $N_{\rm H}\sim1\times10^{21}\,{\rm cm}^{-2}$ warm absorber zones of the \textsc{xstar} grid with ionization parameters ${\rm log}\,\xi\sim3.0$ and ${\rm log}\,\xi\sim1.8$, providing a still poor fit to the data ($\chi^{2}_{\nu}\sim1.5$).

Given the above model of the broad-band data, the remaining residuals lie in the Fe\,K region as can be seen in Figure \ref{fig:FeK_none}. As documented in Pounds et al. (2004) and Lobban et al. (2011) there is evidence of a highly ionized outflow resulting in blue-shifted Fe\,{\rm XXV} and Fe\,{\rm XVVI} absorption lines at $6.79^{+0.07}_{-0.05}$\,keV and $7.12^{+0.03}_{-0.03}$\,keV. A parametrisation of these features with narrow Gaussians yields a good fit to the data with $\chi^{2}_{\nu}\sim1.1$, see Table \ref{tab:parametrisation}. 

Replacing the \textsc{pexrav} with the \textsc{reflionx} reflection model results in a poor fit to the reflection spectrum, predominantly in the HXD and BAT. In order to supplement the reflection continuum, we add a high column partial coverer (Table \ref{tab:baseline}) which improves the fit by $\triangle\chi^{2}=-62$ (as above in MCG--6-30-15 and NGC 3516, no such excesses remain when using \textsc{pexrav}, $R=1.52^{+0.26}_{-0.29}$), resulting in a good fit to both the high energy data and the reflection continuum. We find a sub-solar iron abundance of $Z_{\rm Fe}=0.5^{+0.4}_{-0.4}$ consistent with Lobban et al. (2011). We also find an emission feature at $2.22^{+0.03}_{-0.03}$\,keV with an improvement of $\triangle\chi^{2}=-16$ which may be due to 1s-3p Si\,{\rm XIII}. 

Lobban et al. (2011) find that the high ionization absorption lines in the Fe\,K region are best fit by a highly ionized \textsc{xstar} grid with a turbulent velocity of $3000\,{\rm km}\,{\rm s}^{-1}$, however, in this analysis we continue to use a turbulent velocity of only $1000\,{\rm km}\,{\rm s}^{-1}$ to remain consistent with the procedure used above and outlined in Section 3.1.4. This still models the absorption features well, improving the fit by a further $\triangle\chi^{2}=-77$ with parameters consistent with Lobban et al. (2011). Subsequent to modelling the highly ionized zone and narrow Fe\,{\rm XXV} emission, there are very few residuals remaining below 6.4\,keV which indicates that relativistic emission from the inner regions of the accretion disc may not be required in NGC 4051 (see Figure \ref{fig:baseline}). However, we also test for the possibility of a relativistic component without the use of a partial covering component in Section 3.3.1.

\begin{table*}
\caption{\textsc{kerrdisk} model parameters in addition to the baseline model. Absorption and \textsc{compTT} components are consistent with those quoted in Table \ref{tab:baseline} and are not stated again here to avoid repetition. * Denotes a frozen parameter. $^{a}$ Flux given in units $(10^{-5}\,{\rm ph\,cm^{-2}\,s^{-1}})$. * Denotes a frozen parameter. A positive $\triangle\chi^{2}$ represents a worsening in the fit upon the removal of the component. Spin cannot be constrained in NGC 3516 at the 90\% confidence level.}
\begin{tabular}{l c c c c c} 
\hline
& Fairall 9 & MCG--6-30-15 & NGC 3516 & NGC 3783 & NGC 4051 \\
\hline
\multicolumn{6}{c}{Kerrdisk Profile} \\
LineE (keV) & 6.4* & 6.4* & 6.4* & 6.4* & 6.4* \\ [0.8ex]
$EW$ (eV) & $89^{+12}_{-17}$ & $161^{+46}_{-44}$ & $46^{+16}_{-36}$ & $45^{+17}_{-26}$ & $156^{+15}_{-54}$ \\ [0.8ex]
$q$ & $3.3^{+2.6}_{-0.4}$ & $2.7^{+0.2}_{-0.1}$ & $2.5^{+1.5}_{-1.0}$ & $2.6^{+0.2}_{-0.1}$ & 3.0* \\ [0.8ex]
$a$ & $0.67^{+0.10}_{-0.11}$ & $0.49^{+0.20}_{-0.12}$ & -- & $<0.32$ & $<0.94$\\ [0.8ex]
$i^{\circ}$ & $33^{+3}_{-3}$ & $44^{+6}_{-2}$ & $51^{+11}_{-5}$ & $<17$ & $46^{+3}_{-16}$\\ [0.8ex]
Flux$^{a}$ & $3.12^{+0.41}_{-0.60}$ & $5.73^{+1.63}_{-1.57}$ & $1.00^{+0.35}_{-0.36}$ & $3.83^{+1.12}_{-0.72}$ & $3.45^{+0.33}_{-1.20}$ \\ [0.8ex]
$\triangle\chi^{2}$ ($a=0.988$) & 6 & 70 & 1 & 15 & 6 \\ [0.8ex]
$\triangle\chi^{2}$ ($a=0$) & 8 & 10 & 1 & 0 & 3 \\ [0.8ex]
$\triangle\chi^{2}$ (without \textsc{kerrdisk}) & 35 & 59 & 10 & 32 & 3 \\
\hline
Fe\,{\rm XXV} emission & \checkmark & -- & -- & -- & -- \\
\hline
Fe\,{\rm XXVI} emission & \checkmark & \checkmark & -- & \checkmark & -- \\
\hline
$\chi^{2}_{\nu}$ & 924.4/881 & 1967.8/1819 & 519.9/497 & 1463.0/1374 & 1088.2/1083 \\
\hline
\end{tabular}
\label{tab:kerrdisk}
\end{table*}

\begin{figure*}
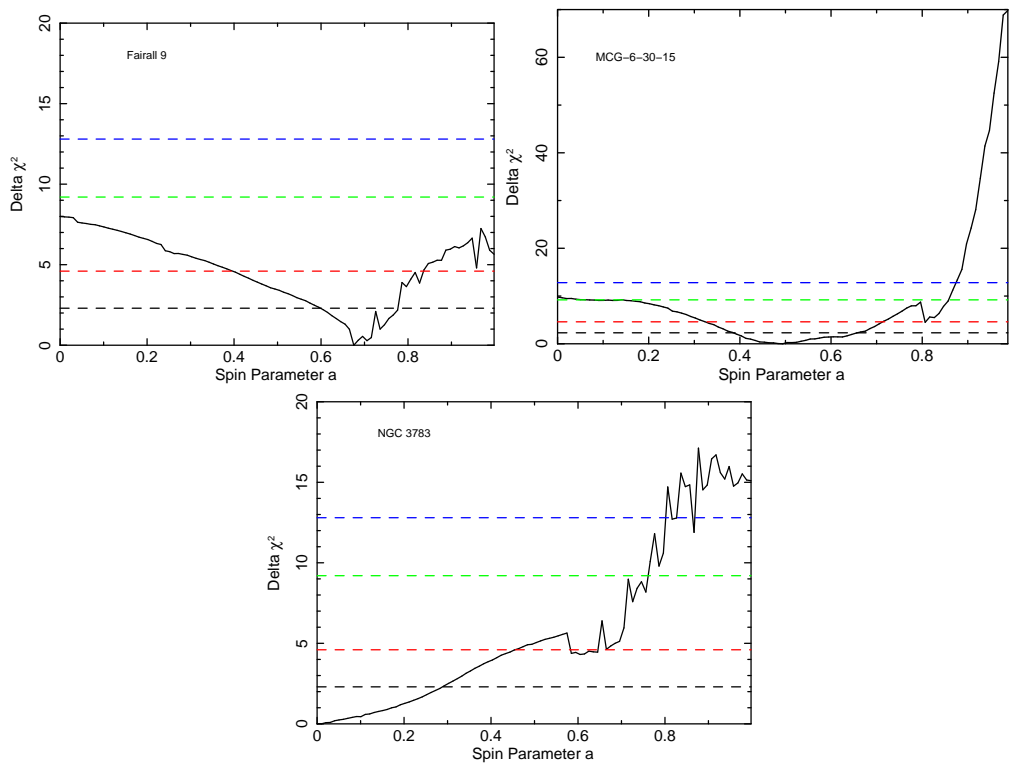

\rotatebox{-90}{\includegraphics[width=5cm]{fairall9_long_99it_paper.ps}}
\rotatebox{-90}{\includegraphics[width=5cm]{mcg6_99it_paper.ps}}
\rotatebox{-90}{\includegraphics[width=5cm]{ngc3783_99it_paper.ps}}
\caption{Fit statistic as a function of $a$ plots for spin constraints made with the baseline+\textsc{kerrdisk} models. Dashed lines representing 68\% (black), 90\% (red), 99\% (green) and 99.9\% (blue) confidence levels for two interesting parameters. Note that the emissivity index q is a free parameter.}
\label{fig:baseline_spin}
\end{figure*}

\begin{figure*}
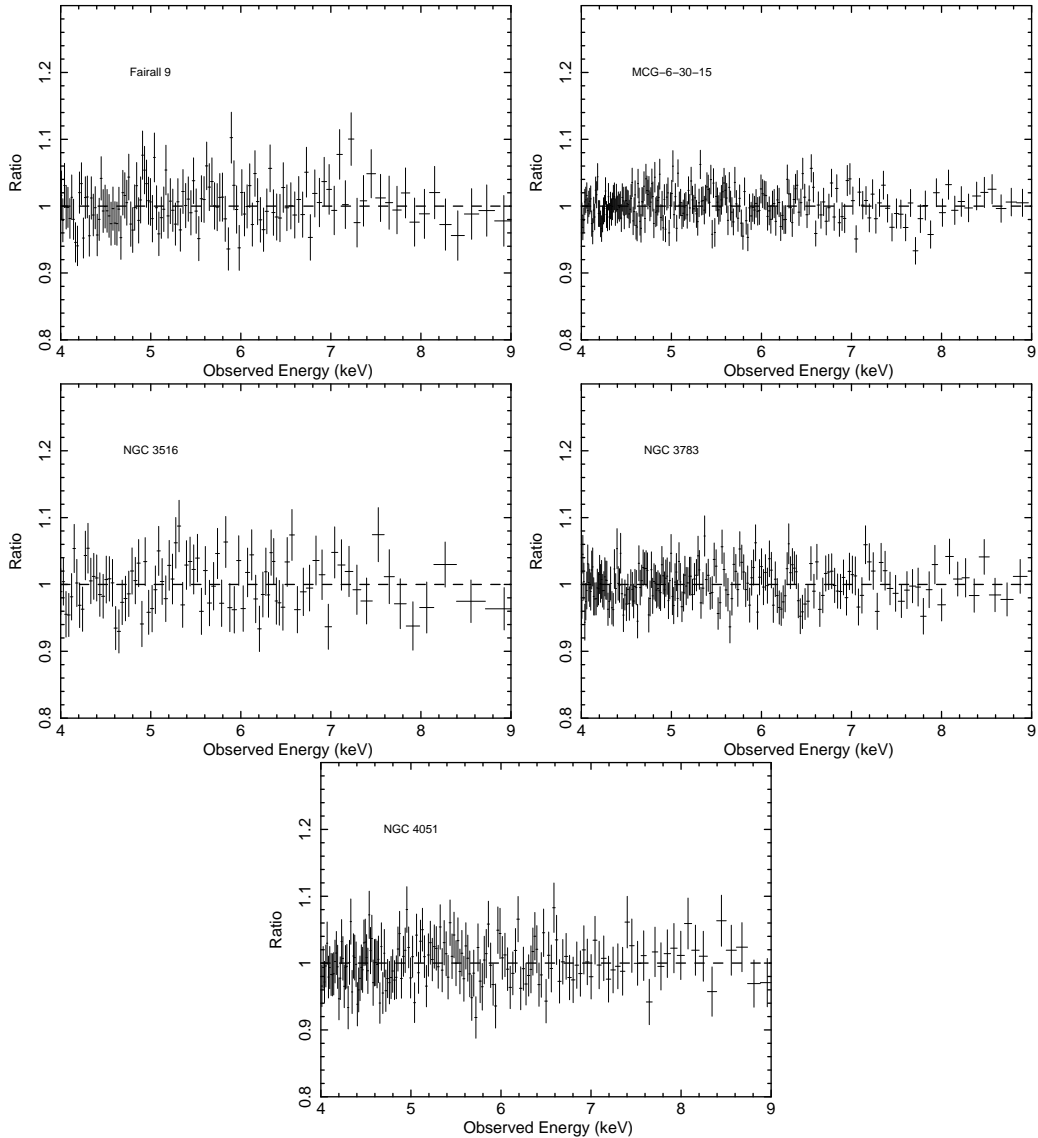

\rotatebox{-90}{\includegraphics[width=5cm]{fairall9_kerrdisk_ratio.ps}}
\rotatebox{-90}{\includegraphics[width=5cm]{mcg6_kerrdisk_ratio.ps}}
\rotatebox{-90}{\includegraphics[width=5cm]{ngc3516_kerrdisk_ratio.ps}}
\rotatebox{-90}{\includegraphics[width=5cm]{ngc3783_kerrdisk_ratio.ps}}
\rotatebox{-90}{\includegraphics[width=5cm]{ngc4051_kerrdisk_ratio_rebin.ps}}
\caption{Ratio plots of the baseline+\textsc{kerrdisk} model between 4-9\,keV. After the application of the \textsc{kerrdisk} relativistic line emission model there are few remaining residuals below 6.4\,keV, giving a very good fit to the Fe\,K region and the broad-band data as a whole.}
\label{fig:kerrdisk_ratio}
\end{figure*}

\begin{figure*}
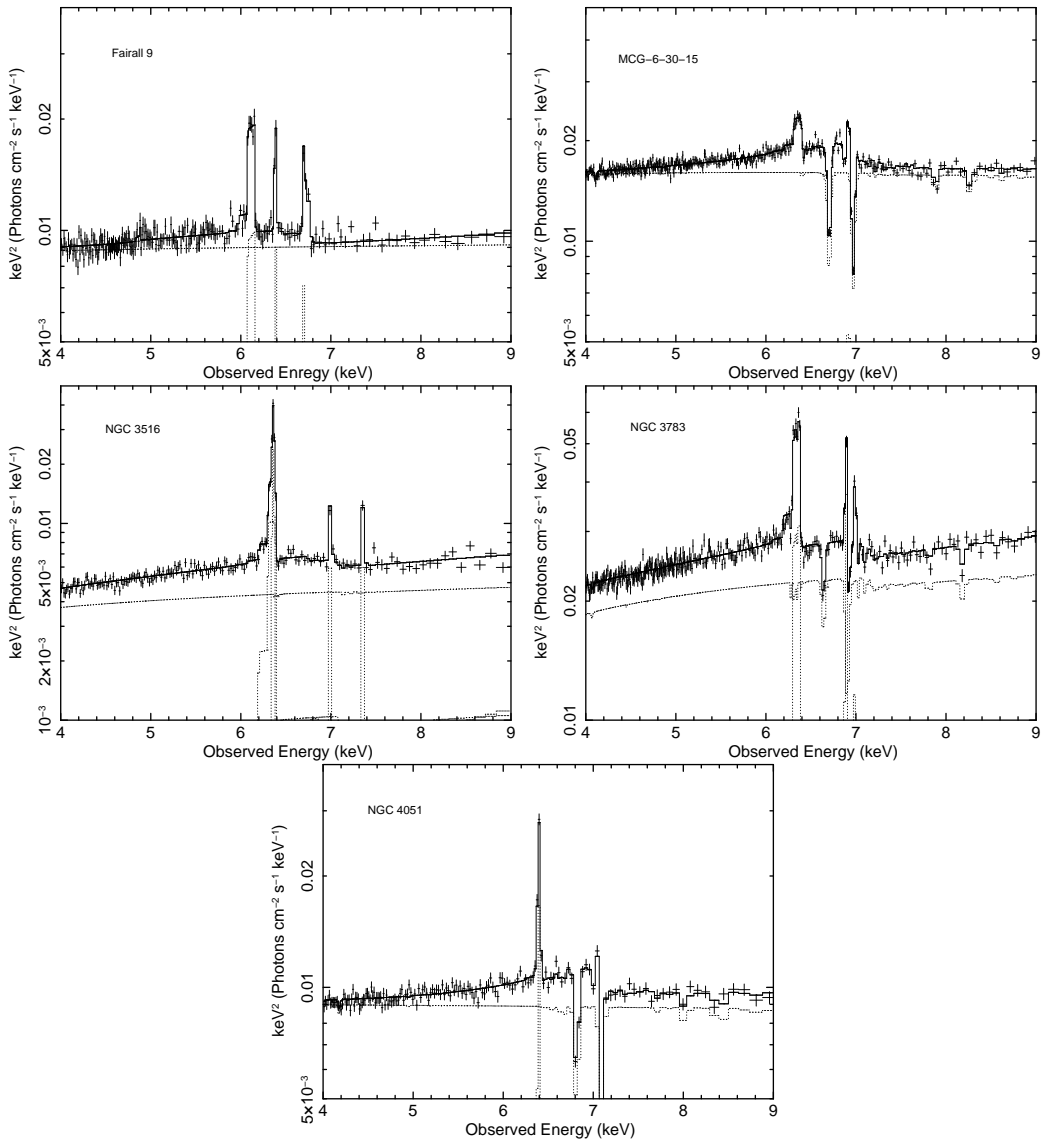

\rotatebox{-90}{\includegraphics[width=5cm]{fairall9_kerrdisk_eeuf.ps}}
\rotatebox{-90}{\includegraphics[width=5cm]{mcg6_kerrdisk_eeuf.ps}}
\rotatebox{-90}{\includegraphics[width=5cm]{ngc3516_kerrdisk_eeuf.ps}}
\rotatebox{-90}{\includegraphics[width=5cm]{ngc3783_kerrdisk_eeuf.ps}}
\rotatebox{-90}{\includegraphics[width=5cm]{ngc4051_kerrdisk_eeuf_rebin.ps}}
\caption{$\nu$F$\nu$ plots of the baseline+\textsc{kerrdisk} model in the 4-9\,keV region. Note the strong Fe\,K band absorption in MCG--6-30-15, NGC 3783 and NGC 4051.}
\label{fig:kerrdisk_eeuf}
\end{figure*}

\subsection{Kerrdisk profile}
Given the above baseline model in Table \ref{tab:baseline} for each object, we can accurately examine any residuals in the Fe\,K region. As a step towards assessing the strength of any contribution from relativistically blurred emission from the inner regions of the accretion disc, we employ the \textsc{kerrdisk} line profile model with BH spin as a variable parameter (Brenneman \& Reynolds 2006). Of course, this is under the assumption that the broad residuals remaining in the Fe\,K region are due to emission from the inner regions of the accretion disc.

This allows us to measure the strength of the relativistic emission, the inner radius of the disc, the inclination and the emissivity of the disc. The emissivity index scales as $R^{-q}$ where $q\sim3$ would be expected from the inner regions of an accretion disc and $q>5$ indicates that emission from the accretion disc is very centrally concentrated with a significant amount of light bending. We assume the accretion disc extends down to the inner-most stable circular orbit ($R_{\rm ISCO}$), which itself varies according to the value of the BH spin parameter. We fix $R_{\rm out}=400\,R_{\rm ms}$ and assume relativistically blurred neutral Fe\,K$\alpha$ with rest frame energy 6.4\,keV. The \textsc{kerrdisk} model only considers the possibility of prograde BH spin with $0<a<0.998$ where $a=cJ/GM^2$ ($J$ is angular momentum). 

It should be noted that the results obtained here can only be judged as estimates due to the possible model degeneracies. In the event that residuals in the Fe\,K region result from the observation of relativistically blurred diskline emission, we attempt to consider these degeneracies and present our best estimates of the line parameters. When employing a diskline model (e.g. \textsc{kerrdisk}), care should be taken due to the interplay between the emissivity index $q$ and the spin parameter $a$. As such the spin versus $\triangle\chi^{2}$ plots in Figure \ref{fig:baseline_spin} indicate the confidence levels for two interesting parameters where both $q$ and $a$ are allowed to vary. The presence of narrow ionized emission lines in the Fe\,K region can also affect the measured emissivity index $q$ and hence the value of the BH spin obtained with a diskline model. However, in these deep observations the number of counts is generally sufficient to distinguish between the presence of a narrow Gaussian or the blue-wing of a relativistically blurred line profile. 

%

The analysis of the long 2010 {\sl Suzaku} Fairall 9 observation here yields, similar results to those of the shorter 2007 observation in Patrick et al. (2011). The disc emissivity is consistent with $q=3.3^{+2.6}_{-0.4}$ although despite the higher signal to noise data, it is not as well constrained (compared to $q=2.7^{+0.7}_{-0.4}$). In an analysis by Schmoll et al. of the same 2007 data they place only a lower limit of $q>4.9$. A significant difference which is responsible for the poorly constrained emissivity index is the clearly detected narrow Fe\,{\rm XXV} emission which is now included in the fit to the Fe\,K region. The interplay between the narrow Gaussian used to model the feature at $\sim6.7$\,keV and the blue-wing of the \textsc{kerrdisk} line profile causes degeneracy between the emissivity index and Fe\,{\rm XXV} flux. The inclusion of narrow Fe\,{\rm XXV} emission also results in a change to the inferred value of the spin parameter $a$, here it is estimated to be $a=0.67^{+0.10}_{-0.11}$, whereas a lower value of $a=0.44^{+0.04}_{-0.11}$ is obtained in the previous analysis when the narrow emission was not included (Figure \ref{fig:baseline_spin}). The findings made here are, however, consistent with an analysis of a 130\,ks {\sl XMM-Newton} observation of Fairall 9 in 2009 by Emmanoulopoulos et al. (2011) who find $a=0.39^{+0.48}_{-0.30}$.

Maximal spin is ruled out in all objects, including MCG--6-30-15 in contrast to claims to the contrary by Reynolds et al. (2005) and Miniutti et al. (2007) of $a>0.93$ and $a>0.92$ respectively. One possible difference is that a full broad-band model has been compiled to fit the 0.6-100.0\,keV spectrum, whereas the Miniutti et al. (2007) spin measurement is only valid over the 3-45\,keV range and without including warm absorbers. In this analysis we find a fairly typical emissivity index of $q=2.7^{+0.2}_{-0.1}$, firstly suggesting that emission in MCG--6-30-15 is not particularly centrally concentrated and the effects of light bending may be minimal. The spin of the central black hole in MCG--6-30-15 is estimated to a much lower spin $a=0.49^{+0.20}_{-0.12}$ (Table \ref{tab:kerrdisk}), maximal spin is ruled out at $>99.5\%$ confidence in this model where the fit is made worse by $\triangle\chi^{2}=+70$ for the case when the spin parameter is fixed at $a=0.998$ (see Figure \ref{fig:baseline_spin}).

\begin{table}
\caption{\textsc{kerrdisk} model parameters when the reflection continuum is modelled with \textsc{pexrav} in place of the \textsc{reflionx} plus partial coverer configuration i.e. no partial covering component is used here. $^{a}$ Flux given in units $(10^{-5}\,{\rm ph\,cm^{-2}\,s^{-1}})$. * Denotes a frozen parameter. A positive $\triangle\chi^{2}$ represents a worsening in the fit. Spin cannot be constrained in NGC 3516 at the 90\% confidence level.}
\begin{tabular}{l c c c c} 
\hline
& MCG--6-30-15 & NGC 3516 & NGC 4051 \\
\hline
& \multicolumn{3}{c}{Kerrdisk Profile} \\
LineE (keV) & 6.4* & 6.4* & 6.4* \\ [0.8ex]
$EW$ (eV) & $264^{+22}_{-27}$ & $88^{+13}_{-26}$ & $88^{+23}_{-25}$ \\ [0.8ex]
$q$ & $2.7^{+0.2}_{-0.2}$ & $<2.7$ & 3.0* \\ [0.8ex]
$a$ & $0.82^{+0.01}_{-0.09}$ & -- & $<0.81$ \\ [0.8ex]
$i^{\circ}$ & $45^{+1}_{-1}$ & $31^{+3}_{-2}$ & $34^{+4}_{-4}$ \\ [0.8ex]
Flux$^{a}$ & $9.01^{+0.73}_{-0.93}$ & $1.87^{+0.28}_{-0.56}$ & $1.95^{+0.51}_{-0.55}$ \\ [0.8ex]
$\triangle\chi^{2}$ ($a=0.998$) & 17 & 12 & 10 \\ [0.8ex]
$\triangle\chi^{2}$ ($a=0$) & 39 & 2 & 2 \\ [0.8ex]
$\triangle\chi^{2}$ (without) & 354 & 35 & 28 \\
\hline
Fe\,{\rm XXV} emission & -- & -- & -- \\
\hline
Fe\,{\rm XXVI} emission & \checkmark & -- & -- \\
\hline
$\chi^{2}_{\nu}$ & 2008.8/1821 & 536.9/498 & 1131.2/1086 \\
\hline
\end{tabular}
\label{tab:pexrav_kerrdisk}
\end{table}

\begin{figure}
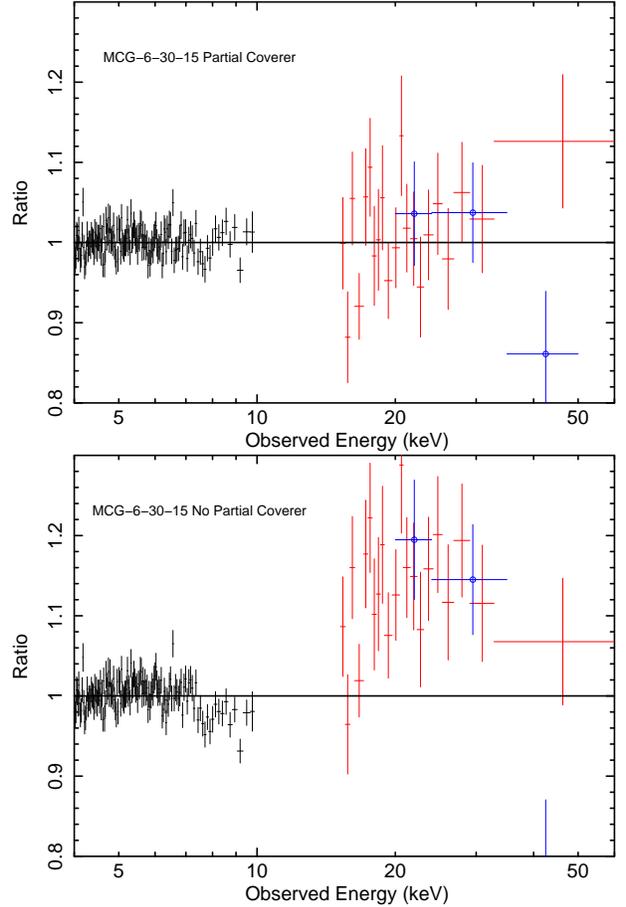

\rotatebox{-90}{\includegraphics[width=6cm]{mcg6_kerrdisk_pc.ps}}
\rotatebox{-90}{\includegraphics[width=6cm]{mcg6_kerrdisk_no_pc.ps}}
\caption{Comparison between the fit to the 4-60\,keV data for MCG--6-30-15 with and without a partial covering component. The first plot shows the baseline model, in the second plot the partial coverer has been removed and the data refit i.e. still using \textsc{reflionx} and not \textsc{pexrav}. Note that the (high column) partial covering component therefore affects predominantly the hard X-ray data, with little affect upon the Fe\,K region. HXD data is in red and BAT data from {\sl Swift} is represented by blue circles.}
\label{fig:pc_no_pc}
\end{figure}

\begin{table}
\centering
\caption{\textsc{relline} model parameters in NGC 3783 allowing for retrograde SMBH spin. $^{a}$ Flux given in units $(10^{-5}\,{\rm ph\,cm^{-2}\,s^{-1}})$. * Denotes a frozen parameter. A positive $\triangle\chi^{2}$ represents a worsening in the fit.}
\begin{tabular}{l c c} 
\hline
& NGC 3783  \\
\hline
& Relline Profile \\
LineE (keV) & 6.4* \\ [0.8ex]
$EW$ (eV) & $43^{+8}_{-16}$\\ [0.8ex]
$q$ & $3.2^{+0.4}_{-0.4}$ \\ [0.8ex]
$a$ & $<-0.04$ \\ [0.8ex]
$i^{\circ}$ & $<17$ \\ [0.8ex]
Flux$^{a}$ & $3.94^{+0.73}_{-1.46}$ \\ [0.8ex]
$\triangle\chi^{2}$ ($a=0.998$) & 13 \\ [0.8ex]
$\triangle\chi^{2}$ ($a=0$) & 3 \\ [0.8ex]
$\triangle\chi^{2}$ (without \textsc{relline}) & 44 \\
\hline
Fe\,{\rm XXV} emission & -- \\
\hline
Fe\,{\rm XXVI} emission & \checkmark \\
\hline
$\chi^{2}_{\nu}$ & 1452.5/1374 \\
\hline
\end{tabular}
\label{tab:ngc3783_relline}
\end{table}

Spin cannot be constrained in NGC 3516 and only loose constraints can be made on the emissivity index of the disc $q=2.5^{+1.5}_{-1.0}$, this is expected due to the relatively weak line modelled using a broad Gaussian in Table \ref{tab:parametrisation}. The fit is only improved by $\triangle\chi^{2}=-10$ after adding \textsc{kerrdisk} emission with $EW=46^{+16}_{-36}$\,eV and no further narrow ionized emission is required in the Fe\,K region with the \textsc{kerrdisk} profile taking the place of the narrow Fe\,{\rm XXV} emission (see Figure \ref{fig:kerrdisk_ratio}). However, the Fe\,K region is well described by the baseline model (Table \ref{tab:baseline}) which includes narrow Fe\,{\rm XXV} emission, adding a \textsc{kerrdisk} line profile only improves the fit in this case by $\triangle\chi^{2}=-2$ over the baseline model.

Only an upper limit of $a<0.32$ at the 90\% confidence level can be placed upon the spin in NGC 3783 using the \textsc{kerrdisk} model. A maximally rotating BH is again ruled out with a worsening to the fit of $\triangle\chi^{2}=+15$. Similarily only an upper limit can be placed upon the inclination of the accretion disc with $i<17^{\circ}$, however, the emissivity can be tightly constrained to $q=2.6^{+0.2}_{-0.1}$. The strength of the broad line component is consistent with Reeves et al. (2004) ($EW=46^{+16}_{-36}$\,eV here compared to $EW=58^{+12}_{-12}$\,eV), being reduced significantly after including a physical model for the highly ionized absorption. 

The broad line feature in NGC 4051 makes a statistically insignificant improvement to the quality of the fit with a reduction of $\triangle\chi^{2}=-3$. Hence, none of the accretion disc parameters can be particularly well constrained, the emissivity index is fixed at $q=3$ and only an upper limit can be made upon the spin parameter of $a<0.94$. 
The Fe\,{\rm XXV} emission listed in Table \ref{tab:baseline} is no longer required with the introduction of the \textsc{kerrdisk} component and is alternatively explained as a blue-wing to the line profile (Figure \ref{fig:kerrdisk_eeuf}). The results obtained here using a highly ionized \textsc{xstar} grid with a turbulent velocity of $1000\,{\rm km}\,{\rm s}^{-1}$ are consistent with those when a turbulent velocity of $3000\,{\rm km}\,{\rm s}^{-1}$ is used (e.g. by Lobban et al. 2011) i.e. line strength, $EW$ and $R_{\rm in}$. 

\subsubsection{The effect of partial covering}
As stated in Section 3.2.2, the high column density partial covering component affects predominantly the hard X-ray data in the HXD and BAT detectors, however, we test the effect of this partial covering component on any curvature in the Fe\,K region. The partial covering component and distant reflection from \textsc{reflionx} are removed and replaced with the \textsc{pexrav} neutral reflection model (plus a narrow neutral 6.4\,keV Gaussian) which achieved a good fit to the data in the initial parametrisation. This interpretation i.e. without partial covering, yields a slightly higher spin constraint of $a=0.82^{+0.01}_{-0.09}$ for MCG--6-30-15, however, a maximally rotating BH is still discounted at an increase of $\triangle\chi^{2}=+17$ ($>99.5\%$ confidence) despite a fairly low reflection fraction of $R=0.77^{+0.10}_{-0.11}$. Figure \ref{fig:pc_no_pc} shows the residuals to the data with the partial covering component removed (still including \textsc{reflionx}), showing that the high column partial covering component in MCG--6-30-15 predominantly affects the fit to the hard X-ray data. Only subtle curvature is introduced to the 7-10\,keV spectrum in the XIS detectors. 

Both NGC 3516 and NGC 4051 (which display no strong broadened emission) also feature a partial covering component in the baseline model, removing the partial coverer and instead modelling the hard excess using \textsc{pexrav} gives accretion disc parameters which are entirely consistent with those in Table \ref{tab:kerrdisk} (see Table \ref{tab:pexrav_kerrdisk}). This again suggests that significant relativistically broadened emission is probably not strongly present in these objects and the high column partial coverer does not have a significant affect upon the Fe\,K modelling. The partial covering component as used in the baseline model does have some affect upon the derived accretion disc and BH parameters, although only in MCG--6-30-15, for example the measured strength (i.e. flux and $EW$) is slightly higher without the presence of a partial coverer as is the measured BH spin (see Table \ref{tab:pexrav_kerrdisk}). Properties of the accretion disc such as the emissivity index and the inclination remain relatively unchanged. Despite the removal of the partial covering component in all objects (where present), a maximally spinning central BH is ruled out in these objects. Note that this is not to say that the residuals in the Fe\,K region cannot be accounted for by a partial coverer, simply that the high column partial coverer used here does not significantly affect the measured accretion disc parameters.

\subsubsection{Testing for retrograde SMBH spin}
Recent work has suggested the possibility of retrograde SMBH spin in AGN, whereby the rotation of the central BH and the surrounding accretion disc are anti-parallel whilst maintaining a stable configuration (King et al. 2005). The \textsc{relline} model (Dauser et al. 2010) is a relativistic line emission model accomodating both negatively and positively spinning central BHs, which produces line profiles consistent with established prograde spin line emission models e.g. \textsc{kerrdisk} (Brenneman \& Reynolds 2006) and \textsc{kyrline} (Dov\v{c}iak et al. 2004). 

The spin versus $\triangle\chi^{2}$ plot for NGC 3783 in Figure \ref{fig:baseline_spin} reveals a best-fitting spin parameter of $a=0$ according the \textsc{kerrdisk} model, assuming a strictly prograde BH spin, indeed NGC 3783 is the only object showing this trend in this sample. We use the \textsc{relline} model in addition to the baseline model to assess the suitability of a retrograde BH in a fit to the Fe\,K region in NGC 3783. In a similar procedure to that employed using the \textsc{kerrdisk} model, we assume the accretion disc extends down to $R_{\rm ISCO}$ and that emission occurs from neutral Fe\,K$\alpha$ and as such we fix the line energy at 6.4\,keV. 

\begin{figure}
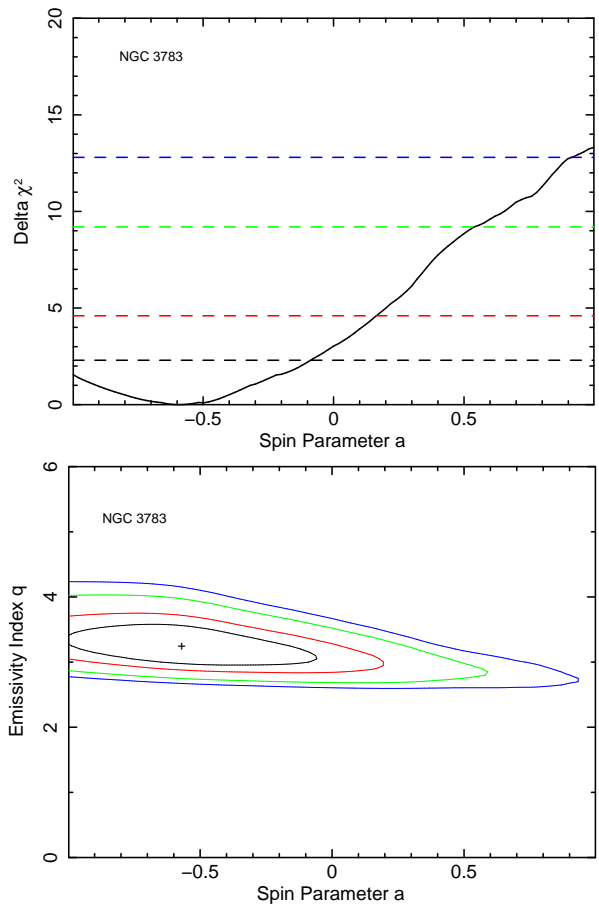

\rotatebox{-90}{\includegraphics[width=6cm]{ngc3783_relline.ps}}
\rotatebox{-90}{\includegraphics[width=6cm]{2D_relline.ps}}
\caption{The upper plot is a fit statistic versus $a$ plot for SMBH spin from the \textsc{relline} model when used in conjunction with the baseline model in NGC 3783. The spin parameter $a$ is allowed to cover the full range $-0.988<a<+0.998$ to assess the possibility of retrograde BH spin. The lower plot shows a 2D contour plot showing the relation between emissivity index and spin, confidence levels labelled in both plots are 68\% (black), 90\% (red), 99\% (green) and 99.9\% (blue) for two interesting parameters.}
\label{fig:ngc3783_relline}
\end{figure}

The accretion disc parameters obtained are very similar to those obtained with \textsc{kerrdisk} in Table \ref{tab:kerrdisk} with a slightly higher emissivity of $q=3.2^{+0.4}_{-0.4}$, inclination $i<17^{\circ}$ and $EW=43^{+8}_{-16}$\,eV (see Table \ref{tab:ngc3783_relline}). Modelling the Fe\,K region of NGC 3783 allowing for retrograde spin suggests an upper limit of $a<-0.04$, as shown in Figure \ref{fig:ngc3783_relline}. This modelling of NGC 3783, therefore, suggests a negatively spinning BH in NGC 3783 at the 90\% confidence level, ruling out a maximally spinning {\sl prograde} BH at $>99.5\%$ confidence with an increase of $\triangle\chi^{2}\sim+13$ in fit statistic when fixed at $a=+0.998$. Figure \ref{fig:ngc3783_relline} also shows a 2D emissivity vs spin contour plot for NGC 3783 using \textsc{relline}, indicating the relation between these two parameters. It is interesting to note that the emissivity of the accretion disc is well constrained and, in this object, has little effect upon the measured BH spin. There is only a slight negative correlation in which emission is more centrally concentrated for lower (more negative) BH spin. 

It should be noted that while a retrograde BH does indeed push the innermost stable circular orbit of the accretion disc outwards, thereby reducing the extent to which the line profile can become relativistically broadened, the line profile is noticibly broadened and not simply representative of a series of narrow Gaussians which would otherwise be indicative of purely distant emission (see Figure \ref{fig:relline_all}).

\subsection{Dual reflector}
The previous \textsc{kerrdisk} relativistic line emission model is representative of reflection off the inner regions of the accretion disc. If strong emission arises from this region, the same relativistic broadening must also be applied to the reflection continuum as a whole, as opposed to purely the narrow 6.4\,keV Fe\,K$\alpha$ emission. To account for this we replace the \textsc{kerrdisk} model with a second \textsc{reflionx} component convolved with the \textsc{kerrconv} kernel (Brenneman \& Reynolds 2006) in addition to the unblurred reflection component. We also remove the partial covering component from the dual reflector fits to assess this alternative interpretation. 

In some analyses of AGN, such a blurred reflection component has been used to model the soft excess under the assumption that the regions responsible for the smearing of discrete atomic emission features at soft energies are also responsible for broadening in the Fe\,K region (e.g. Schmoll et al. 2009; Nardini et al. 2010). However, in an albeit small sample of bare Seyfert AGN, Patrick et al. (2011) derived contrasting accretion disc parameters depending upon whether such blurring is allowed to model the soft excess in addition to features in the Fe\,K region. A significant amount of blurring is required in order to smear the soft atomic emission features into a smooth soft excess, resulting in much higher derived BH spin and very centrally concentrated emission (i.e. high $q>4$). 

Taking this into account, in this interpretation of the data, we retain the use of the \textsc{compTT} Comptonization origin to model the soft excess, leaving the convolved reflection spectrum to account for any observed iron line broadening and hard excess over purely distant reflection. A purely reflection origin for the hard excess is assumed and no partial covering model is used in the dual reflector fit as in MCG--6-30-15, NGC 3516 and NGC 4051. The iron abundance of both the inner and distant reflector are tied and frozen at the best fitting values from the baseline model through the fits. 

A good fit can be obtained to all objects with a dual reflector model (Table \ref{tab:dual_reflector}), with the exception of MCG--6-30-15, the hard X-ray data from the HXD and BAT detectors are fit well without the requirement for any additional reflection or a partial covering component. Due to the increased number of free model parameters and hence complexity in the broadband model it is difficult to measure accretion disc parameters (particularly BH spin) to any reasonable accuracy, even with high quality data as used here from deep {\sl Suzaku} observations. Consequently, no better constraints can be made using a dual reflector fit, rather the results obtained are generally consistent with the baseline+\textsc{kerrdisk} model (Table \ref{tab:kerrdisk}).

For the case of MCG--6-30-15, a reasonable quality of fit ($\chi^{2}_{\nu}=2061.4/1823$) is obtained with a dual reflector fit. As noted above, the interpretation of the soft excess can dramatically alter the measured accretion disc parameters. In an object as complex as MCG--6-30-15 with strong reflection, broadened relativistic emission in the Fe\,K region, complex warm absorption and a soft excess, the use of a dual reflector results in a number of degeneracies when modelling the spectrum. As a result of this we fix the \textsc{compTT} parameters at the best-fit values obtained with the baseline model and then proceed to add a blurred reflection component in place of the partial covering model. This approach ensures that the blurred reflection component does not model the soft excess in place of \textsc{compTT} and as such the accretion disc emissivity and BH spin will not be {\sl forced} to extreme values in order to reproduce a smooth and featureless continuum.

The measured parameters for MCG--6-30-15 are consistent with those obtained in Table \ref{tab:kerrdisk} with $q=2.3^{+0.2}_{-0.1}$ and $a=0.61^{+0.15}_{-0.17}$. Also requiring a Fe\,{\rm XXV} emission line which improves the fit by $\triangle\chi^{2}=-7$, however it is weak with a low $EW=6^{+5}_{-3}$\,eV. Solar abundances are assumed throughout and a consistent ionization parameter ($\xi<11$\,erg\,cm\,s$^{-1}$) is measured in both the blurred and unblurred reflectors. The warm absorber parameters measured with the dual reflector fit are also consistent with those obtained in the baseline model.


The properties of the accretion disc are not greatly affected by the interpretation of the origin of the hard excess i.e. whether a high column ($N_{\rm H}\sim10^{24}\,{\rm cm}^{-2}$) partial covering model or a dual blurred and unblurred reflector is used (see Table \ref{tab:dual_reflector}). The partial covering model used in the baseline model therefore has a limited effect upon the accretion disc parameters and BH spin measured from the Fe\,K region. The results obtained with the dual reflector fit in NGC 3783 are completely consistent with those measured with the previous baseline+\textsc{kerrdisk} model, again an upper limit on the BH spin parameter is found at $a<0.31$ and the emissivity of the accretion disc is also a similar $q=2.8^{+0.4}_{-0.2}$. 

The complex absorption in these objects (other than Fairall 9) can mask the extent to which the smeared discrete emission features are present in the data at soft X-ray energies. More extreme blurring (a high emissivity, for example) is not required since the X-ray spectrum at soft energies in these more complex objects is far from smooth and the discrete emission features do not require smearing into a smooth continuum-like shape. As a consequence of this, the main components driving the dual reflector fits are the features in the Fe\,K region, rather than curvature due to the blurring of soft X-ray atomic emission.

\section{Discussion}
This sample of Seyfert 1 AGN includes deep {\sl Suzaku} observations with exposures $>200$\,ks at high S/N, enabling us to fully model the broad-band spectrum when combined with additional hard X-ray data from {\sl Swift}/BAT. In order to robustly assess the nature of any features in the Fe\,K region, whether it be absorption or relativistically blurred emission, a complete modelling of the X-ray spectrum was undertaken to ensure that all features are taken into account. With complex X-ray spectra such as found in these AGN, deep exposures are essential to help constrain a number of features prior to investigating accretion disc properties and the potential for measuring BH spin.

Some improvement is made to the fit for all objects with the addition of the \textsc{kerrdisk} model which represents blurred line emission from the inner regions of the accretion disc. The overall improvement to the fit for NGC 3516 and NGC 4051 is small with $\triangle\chi^{2}=-10$ and -3 in each for 4 and 3 additional parameters respectively, thus even the presence of disc emission lines in these AGN is very tentative. Statistically much stronger broad line components are found in MCG--6-30-15 and NGC 3783 even after a complete modelling of the warm absorber components, although in some circumstances complex absorption models may be able to account for these features.

Narrow ionized emission is found in three of the objects here after including a \textsc{kerrdisk} line profile, however the blue-wing of the relativistically broadened line profile can easily become degenerate with such emission. For example, the emissivity index and the inclination of the accretion disc can be forced to values such that a narrow feature which would otherwise be attributed to Fe\,{\rm XXV}/ Fe\,{\rm XXVI} emission could be explained by the blue-wing of relativistically blurred diskline emission from neutral Fe\,K$\alpha$. Indeed, NGC 3516 and NGC 4051 require no ionized narrow emission features in the final baseline+\textsc{kerrdisk} since the broad feature can successfully reproduce both the red-wing and narrow-like features in the Fe\,K region, yet there is little requirement for broadened line profiles in either object. Emission from Fe\,{\rm XXV} is required only in Fairall 9 even after the possible degeneracy mentioned above, Fe\,{\rm XXVI} Ly\,$\alpha$ emission is more common and is found in 3/5 objects (Fairall 9, MCG--6-30-15 and NGC 3783). Future calorimeter observations with {\sl Astro-H} (Takahashi et al. 2010) will be able to distinguish and resolve narrow ionized emission from the blue-wing of a diskline profile.

\subsection{Average parameters of the disc}
Given the measured values from the \textsc{kerrdisk} line profile, we can calculate typical parameters of the accretion disc. The average inclination of the disc is $i=37^{\circ}\pm8^{\circ}$, although this is a small sample, it is consistent with Nandra et al. (2007) who find $i=38^{\circ}\pm6^{\circ}$. From the baseline+\textsc{kerrdisk} model (Table \ref{tab:kerrdisk}) we also measure an average accretion disc emissivity of $q=2.8\pm0.2$ which (similarly to Patrick et al. 2011) is a lower value than is often assumed from light bending models (e.g. Gallo et al. 2011; Nardini et al. 2011). The initial parametrisation in Table \ref{tab:parametrisation} with a broad Gaussian measures an average strength of the broad line component of $EW=77\pm23$\,eV at a centroid line energy of $6.11\pm0.10$\,keV, also consistent with the previous findings by Nandra et al. (2007) with {\sl XMM-Newton}. Employing the more physical \textsc{kerrdisk} model increases the strength to $EW=99\pm28$\,eV.

The moderate emissivity indices found here are likely due to the accretion disc parameters being obtained purely from broadening in the Fe\,K region rather than requiring the inner regions of the accretion disc to blur the reflection continuum substantially in order to model the soft excess. Here we assume the soft excess originates from the Comptonization of UV seed photons from the disc in a hot corona above the disc via the \textsc{compTT} model (Titarchuk 1994), a second soft powerlaw or disc blackbody would provide a quantitively similar fit to the data. If discrete atomic features present in the reflection continuum at soft X-ray energies were allowed to model the model the soft excess through relativistic smearing due to originating from regions close to the central BH, the emissivity of the accretion disc would be forced to much higher values i.e. $q>4.5$ (see Patrick et al. 2011). Such high values are not required for a good fit to the Fe\,K region alone, therefore, {\sl if} a blurred reflection component is responsible for the soft excess it must originate from a different part of the disc responsible for broadening in the Fe\,K region. 

The dual reflector fit in Section 3.4 in which two separate \textsc{reflionx} components are used for distant and inner reflection is also consistent with the typical accretion disc measurements made above (Table \ref{tab:dual_reflector}). While higher resolution spectroscopy would be required to fully distinguish between the varying scenarios, the hard excess can be explained by either a high column partial coverer or a dual reflector, although crucially still without the requirement for extreme parameters such as high $q$ and $a$ for very centrally concentrated emission with an average $q=2.4\pm0.2$.

\subsection{The effect of absorbers on the Fe\,K region}
It is known that narrow ionized emission lines from Fe\,{\rm XXV} and Fe\,{\rm XXVI} can alter properties such as the measured inclination and emissivity of the accretion disc, however, the warm absorber is often overlooked. For instance warm absorber components can introduce spectral curvature, due to the bound-free L/\,K-shell transitions of abundant elements (Miller \& Turner 2009). In particular, the warm absorber can play a significant role in the shape of the spectrum, not only at low X-ray energies, but also higher energies closer to the Fe\,K region (Miller et al. 2008, Zycki et al. 2010). Utilising the soft X-ray energies below 3\,keV allows a complete modelling of the warm absorber, adding the corresponding spectral curvature, which may not otherwise be taken into account without using the available data in the 0.6-3.0\,keV range. Given a broad-band model, the true nature of the Fe\,K region and any broad red-wing can be determined after the reduced spectral curvature and observed `broadening' from the required warm absorbers has been taken into account. Inappropriate modelling of the warm absorber can thus alter the observed strength and dominance of the broad red-wing over the continuum, which in turn will affect the extent to which black hole spin is required and its measured value. 

In addition to the warm absorber properties, modelling the highly ionized absorption lines in the Fe\,K region with a physically consistent \textsc{xstar} grid (such as with a turbulent velocity of 1000\,km\,${\rm s}^{-1}$ used here) ensures the corresponding curvature is removed from the spectrum. Highly ionized zones of gas are required in three of the objects here: MCG--6-30-15, NGC 3783 and NGC 4051, also consistent with past studies (Reeves et al. 2004; Miller et al. 2008; Tombesi et al. 2010; Lobban et al. 2011). Absorption lines in the Fe\,K region due to highly ionized species of iron (Fe\,{\rm XXV} and/ or Fe\,{\rm XXVI}) are found at the appropriate energy (or blue-shifted such as in NGC 4051 and to a lesser extent in MCG--6-30-15, with outflow velocities of $5600^{+800}_{-700}$\,${\rm km}\,{\rm s}^{-1}$ and $3200^{+400}_{-500}$\,${\rm km}\,{\rm s}^{-1}$ respectively). NGC 3783 shows both broadened line emission and absorption lines in the Fe\,K region, however when a 1000\,km${\rm s}^{-1}$ high ionization \textsc{xstar} grid is included, the observed strength and shape of the broad line is slightly reduced with \textsc{kerrdisk} $EW=67^{+15}_{-29}$\,eV reducing to $EW=45^{+17}_{-26}$\,eV (although still consistent within errors). The same can be said for the highly ionized absorption present in MCG--6-30-15 with diskline models proving marginally weaker after the introduction of the high ionization \textsc{xstar} grid (i.e. $EW=180^{+29}_{-54}$\,eV reduced to $161^{+46}_{-44}$\,eV).

As discussed in Section 3.3.1, complex partial covering models can provide an alternative explanation of the broadband spectrum and therefore the residuals in the Fe\,K region. For example, Miller et al. (2009) construct a model which can explain all the features in the broadband spectrum of MCG--6-30-15 with no requirement for a diskline component, rather the apparent `broad' residuals in the Fe\,K region are accounted for with `complex clumpy' absorbers. Here, however, we find that even when we apply partial covering absorption models to the X-ray spectra, the addition of a broad iron line component still appears to be required in the case of MCG--6-30-15, but only marginally in NGC 3516 and NGC 4051. 

\subsection{Is MCG--6-30-15 not maximal?}
Previous analyses of MCG--6-30-15 have claimed a very strong, broad iron line whilst requiring that the central black hole is spinning very rapidly in order to account for the extreme width of the broad line. Miniutti et al. (2007) measure $a>0.92$ in a fit which does not consider data below 3\,keV in order to avoid the need to fit for the warm absorber. Here we make use of the full broad-band data available, fitting in the 0.6-100.0\,keV range, using a combination of the XIS, HXD and BAT detectors in order to properly account for any absorption components which may be present in addition to the reflection hump above 10\,keV. Taking the warm absorber components into account prior to considering emission from central regions close to the BH reveals that while the majority of the effects due to the warm absorption occur at low energies ($<3$\,keV), significant spectral curvature is added in MCG--6-30-15 at higher energies (Zycki et al. 2010). Much of the very broad red-wing observed by Miniutti et al. (2007) is reduced, resulting in a much weaker broad line component than has previously been suggested. This results in a broad line feature which may originate from emission close to a BH with spin $a=0.49^{+0.20}_{-0.12}$ and lower emissivity of $q=2.7^{+0.2}_{-0.1}$ while the near maximal spin required by Reynolds et al. (2005) and Miniutti et al. (2007) is excluded at high confidence levels in our model (see Table \ref{tab:kerrdisk} and Figure \ref{fig:baseline_spin}).

In this paper, hard X-ray emission in excess over those which can be reproduced by a single \textsc{reflionx} are accounted for using a high column partial coverer. Indeed in MCG--6-30-15 Miller, Turner \& Reeves (2008) successfully model the entire 0.5-45.0\,keV spectrum through the use of absorbers and partial covering. The full model we form here also includes such a component, although its effect upon the broadening in the Fe\,K region is subtle. However, even if no partial covering model is included, a maximally rotating BH is still ruled out at a high confidence level as can be seen in Table \ref{tab:pexrav_kerrdisk} and Figure \ref{fig:pc_no_pc}.

\begin{figure}
{\includegraphics[width=8.5cm]{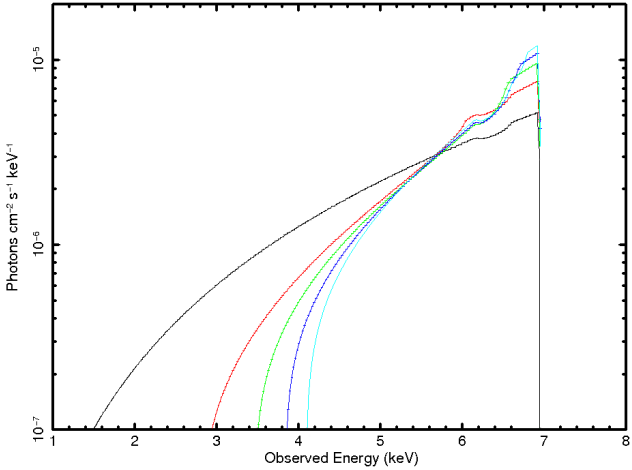}}
\caption{Line profiles using \textsc{relline} with rest energy 6.4\,keV and parameters $q=3.0$, $i=40^{\circ}$ and spin increasing left to right $a=+0.998$ (black), $a=+0.5$ (red), $a=0.0$ (green), $a=-0.5$ (dark blue) and $a=-0.998$ (light blue). Note that even in the case of a maximally rotating retrograde BH (furthest out $R_{\rm ISCO}$), there is still a significantly broadened line profile.}
\label{fig:relline_all}
\end{figure}

\subsection{A retrograde SMBH in NGC 3783?}
Retrograde SMBH spin has been thought of as a mechanism through which the central BH can power radio jets in AGN. Garofalo (2009) notes that accretion onto retrograde black holes can provide a powerful mechanism resulting in higher Blandford-Znajek luminosities than those in the prograde spin case due to $R_{\rm ISCO}$ being located further away from the SMBH, indeed some radio-loud AGN have been found to have inner disc truncations at $\sim8-9\,R_{\rm g}$ e.g. (3C 120, Kataoka et al. 2007) with $R_{\rm ISCO}\sim9\,R_{\rm g}$ for a maximally spinning {\sl retrograde} BH. 

The line profile produced from emission surrounding a negatively spinning BH is narrower than that obtained from a positively spinning BH, however measuring a retrograde spin is not necessarily representative of emission occuring from simply distant regions. The inner radius of emission of the accretion disc varies according to the spin of the central BH with a maximally rotating prograde BH producing $R_{\rm ISCO}=1.23\,R_{\rm g}$, a Schwarzschild BH producing $R_{\rm ISCO}=6\,R_{\rm g}$ and maximally rotating retrograde BH producing $R_{\rm ISCO}\sim9\,R_{\rm g}$ i.e. $R_{\rm ISCO}$ increases with decreasing positive spin. Even assuming a maximally rotating retrograde BH (the largest value of $R_{\rm ISCO}$), emission still occurs close enough to the central black hole for substantial relativistic effects and Doppler broadening. The resultant line profile is not therefore equivalent to purely distant emission. Figure \ref{fig:relline_all} shows a comparison between the line profiles produced with the \textsc{relline} model with spin varying from $a=-0.998$ to $a=+0.998$. However, it could also be that the BH is not retrograde, rather that the inner radius of the accretion disc is truncated short of the ISCO.

In this analysis we find a BH spin in NGC 3783 of $a<-0.04$ at the 90\% confidence level (for one interesting parameter), with Figure \ref{fig:ngc3783_relline} suggesting a best-fitting intermediate retrograde BH spin. Accretion disc parameters obtained with the \textsc{relline} model are consistent with those obtained with the \textsc{kerrdisk} model, measuring an emissivity $q=3.2^{+0.4}_{-0.4}$ and a low inclination of $i<17^{\circ}$. 

A recent analysis of NGC 3783 by Brenneman et al. (2011) of the same {\sl Suzaku} data has suggested a BH spin of $a>0.98$ with an accretion disc emissivity of $q=5.2^{+0.7}_{-0.8}$ using a dual reflector model i.e. \textsc{pexmon}+(\textsc{reflionx*relconv}). This contrasts significantly with the estimates made here, despite Brenneman et al. (2011) including warm absorber components in their fit to the 0.7-45.0\,keV data (while ignoring 1.5-2.5\,keV). Brenneman et al. (2011) also allow for partial covering which is not required in the baseline model formed here (Table \ref{tab:baseline}). The highly ionized absorber in the Fe\,K region detected here (e.g. Fe\,{\rm XXV} absorption, $\triangle\chi^{2}=-33$) and by Reeves et al. (2004) in {\sl XMM-Newton} data is not included in their fit, instead a warm absorber with ${\rm log}\,\xi\sim2.53$ is claimed to model this feature sufficiently. There are also variations in the modelling of the accretion disc itself, for example Brenneman et al. (2011) require a super-solar iron abundance of $Z_{\rm Fe}=3.7^{+0.9}_{-0.9}$ and a broken emissivity law with $q_{\rm in}=5.2^{+0.7}_{-0.8}$, $r_{\rm br}=5.4^{+1.9}_{-0.9}\,R_{\rm g}$ and $q_{\rm out}=2.9^{+0.2}_{-0.2}$, whereas the analysis here provides a good fit to the broad-band data with purely Solar iron abundance and a single (unbroken) emissivity index of $q=3.2^{+0.4}_{-0.4}$ (Table \ref{tab:ngc3783_relline}, using \textsc{relline}, $\chi^{2}_{\nu}=1.057$). 

In order to test our conclusions here, we form a model using essentially the same reflection components i.e. \textsc{pexrav}+K\,$\alpha$+K\,$\beta$+(\textsc{reflionx*relconv}) modified by warm absorption, but maintaining Solar abundances and the single emissivity law. This brief test confirms the results found here and changes the estimated spin to $a=-0.2^{+0.05}_{-0.65}$ ($\chi^{2}_{\nu}=1.072$). Further to this, allowing for a broken emissivity index and allowing the iron abundance to vary does not reproduce the Brenneman et al. (2011) results and relaxes the estimates upon the spin parameter $a$, while the abundance remains approximately Solar and there is no indication for the requirement of a broken emissivity i.e. no significant improvement in the quality of the fit. It is possible that the treatment of the absorbing components, soft excess and assumptions regarding the super-Solar iron abundance and broken emissivity law may lead to the higher spin estimated by Brenneman et al. (2011). The overall fit to the 0.7-45.0\,keV data in Brenneman et al. (2011) is poor, mostly influenced by the soft X-ray energies ($\chi^{2}_{\nu}=1.381$, in contrast to the best-fit model here of $\chi^{2}_{\nu}=1.057$). However, it is not clear whether the narrow soft X-ray emission lines present in the {\sl Suzaku} data are included within the Brenneman et al. (2011) analysis. Indeed a complete modelling of the broad-band data (including any soft emission lines) may alter the BH spin estimate.



\subsection{Summary of {\sl Suzaku} observations so far}
Based upon the results of this paper and those found from {\sl Suzaku} observations of Seyfert 1s with low intrinsic absorption in Patrick et al. (2011), we can begin to draw conclusions regarding the Fe\,K region and the accretion disc and BH properties. We compare only these two samples to ensure full and consistent broad-band models are used and as such estimates of average parameters can be made. The objects we consider are therefore: Ark 120, Fairall 9, MCG-2-14-009, MCG--6-30-15, Mrk 335, NGC 3516, NGC 3783, NGC 4051, NGC 7469 and SWIFT J2127.4+5654. Results are based upon modelling any relativistic broadening in the Fe\,K region using the \textsc{kerrdisk} model i.e. Table \ref{tab:kerrdisk} here and Model D (Table 7) in Patrick et al. (2011). 

We find that a maximally rotating Kerr BH is ruled out in all objects other that NGC 3516 in which no spin constraints can be made at all upon the spin for the 2009 {\sl Suzaku} observation. Intermediate spin constraints can be made for Fairall 9 ($a=0.67^{+0.10}_{-0.11}$), MGC--6-30-15 ($a=0.49^{+0.20}_{-0.12}$), Mrk 335 ($a=0.70^{+0.12}_{-0.01}$), NGC 7469 ($a=0.69^{+0.09}_{-0.09}$) and SWIFT J2127.4+5654 ($a=0.70^{+0.10}_{-0.14}$). The remaining five AGN in this sample have central BHs with a measured spin which is consistent with a non-rotating Schwarzschild BH: Ark 120 ($a<0.94$), MCG-2-14-009 ($a<0.88$), NGC 3516 (no constraints upon $a$), NGC 3783 ($a<0.32$) and NGC 4051 ($a<0.94$). It is possible that a retrogradely spinning BH may be found in some of these objects with an upper limit upon $a$, indeed testing the \textsc{relline} model in an analysis of NGC 3783 has suggested this with $a<-0.04$. This possibility will be investigated in a subsequent paper.

Relativistic emission is statistically formally required in all but NGC 3516 and NGC 4051 i.e. in 80\% of the objects, while such emission in MCG-2-14-009 is only weak with $\triangle\chi^{2}=-12$. Although due to the small size of this sample (10 objects) limits the extent to which we can compare this fraction to larger samples and the Seyfert 1 AGN population as a whole, it is encouraging that Nandra et al. (2007) find broad lines in $\sim70\%$ of objects. However, de la Calle P\'{e}rez et al. (2010) find only $\sim36\%$, albeit in a flux limited sample. 

Over the sample of the 10 objects listed above, we measure average parameters of $q=2.5\pm0.2$, $a=0.65\pm0.05$, $i=37^{\circ}\pm3^{\circ}$ and $EW=115\pm16$\,eV. These compare well to Nandra et al. (2007) who find $i=38^{\circ}\pm6^{\circ}$ and $EW=91.3\pm12.8$\,eV, while de la Calle P\'{e}rez et al. (2010) measure an average emissivity index of $q=2.4\pm0.4$. Although the average parameters obtained here are by no means representative of a larger sample of AGN, each has been modelled in a consistent manner. In future work we aim to expand this analysis to all the Seyfert 1 AGN in the {\sl Suzaku} archive with a similar uniform broad-band analysis.

Other objects in the literature which authors have been able place constraints upon BH spin include Mrk 79 (Gallo et al. 2011) and RBS 1124 (Miniutti et al. 2010). Relativistically broadened emission in the Fe\,K region, although occuring outside $6\,R_{\rm g}$, has also been found in: MCG-5-23-16 (Reeves et al. 2007), MCG+8-11-11 (Bianchi et al. 2010), Mrk 509 (Ponti et al. 2009) and NGC 2992 (Yaqoob et al. 2010); with no broad line component being found in: 1H0419-577 (Turner et al. 2009), CenA (Markowitz et al. 2007), MCG-2-58-22 (Rivers et al. in press), Mrk 766 (Turner et al. 2007), NGC 4258 (Reynolds et al. 2009), NGC 4593 (Markowitz \& Reeves 2009), NGC 5548 (Liu et al. 2010) and NGC 7213 (Lobban et al. 2010).

\section{Conclusions}
Resulting from the work above on this small sample of five deep {\sl Suzaku} observations of Seyfert 1 AGN, we conclude the following:
\begin{enumerate}
\item The warm absorber components (if present) in these AGN significantly influence the Fe\,K region, despite predominantly affecting the X-ray spectrum at lower energies. Fitting the full 0.6-100.0\,keV broadband spectrum ensures the additional spectral curvature due to the warm absorber is accounted for prior to making any conclusions about the accretion disc or central supermassive black hole. Indeed the broad 'red-wing' observed in AGN spectra appears much reduced after a complete and consistent modelling of the soft X-ray spectrum. 

\item There is evidence for absorption features in the Fe\,K region in 3/5 objects in this sample due to highly ionized zones of gas. Absorption due to Fe\,{\rm XXV} is found in MCG--6-30-15, NGC 3783 and NGC 4051 and Fe\,{\rm XXVI} absorption is found in MCG--6-30-15 and NGC 4051. The absorption found in NGC 4051 is blue-shifted (as noted by Lobban et al. 2011) and in MCG--6-30-15 (Young et al. 2005). 

\item Examining the Fe\,K region after modelling the soft X-ray absorber and any highly ionized Fe\,K absorption, we find the initially observed broad component in the data is significantly weaker. It is fitted well by \textsc{kerrdisk} line emission from the inner regions of the accretion disc of mean strength $EW=99\pm28$\,eV. Even in the case of MCG--6-30-15, the strength of the broad line is only $EW=161^{+46}_{-44}$\,eV, compared to a cumulative $EW=320^{+45}_{-45}$\,eV in Miniutti et al. (2007) in which the absorption zones are not fully considered in a fit to the 3-45\,keV spectrum.

\item The use of a high column density ($N_{\rm H}\sim10^{24}\,{\rm cm}^{-2}$) partial covering component to account for the hard excess has only a marginal affect upon the derived accretion disc parameters. Subtle curvature is added to the spectrum at Fe\,K energies due to the partial coverer, however, this is not enough to substantially alter the observed strength of any broad component which may be present. This is particularly notable in MCG--6-30-15, although this is not to say that some complex partial covering models cannot fully account for all the residuals in the Fe\,K region.

\item Using the above conclusions, we estimate an average emissivity index of $q=2.8\pm0.2$ in this small sample of AGN, indicating that emission is not particularly centrally concentrated with the majority of emission from the accretion disc not occurring in regions approaching $R_{\rm ISCO}$. Combining this with the results of Patrick et al. (2011), we find an average emissivity of $q=2.5\pm0.2$ over a sample of 10 objects.

\item This analysis rules out a maximally rotating black hole with the \textsc{kerrdisk} model in Fairall 9, MCG--6-30-15 and NGC 3783 at high confidence levels with increases in $\triangle\chi^{2}$ of +6, +70 and +15 respectively. Fitting without a partial covering component similarly rules out a maximally rotating black hole where the fit is worse by $\triangle\chi^{2}=+24$ in MCG--6-30-15. 

\item Given a full broadband model, including the required absorption components, we estimate a BH spin of $a=0.49^{+0.20}_{-0.12}$ in MCG--6-30-15 at an emissivity of $q=2.7^{+0.2}_{-0.1}$, accretion disc inclination of $i^{\circ}=44^{+6}_{-2}$ and a moderate strength of $EW=161^{+46}_{-44}$\,eV.

\item Similarly we estimate an upper limit of $a<0.32$ and $a<0.94$ for NGC 3783 and NGC 4051 respectively. The spin of the central BH in Fairall 9 is estimated to be $a=0.67^{+0.10}_{-0.11}$ whereas the spin cannot be constrained in NGC 3516.

\item In NGC 3783 we find a marginal indication of retrograde spin, using the \textsc{relline} model yields a retrograde BH spin of $a<-0.04$ at the 90\% confidence level with $q=3.2^{+0.4}_{-0.4}$.
\end{enumerate}

\section*{Acknowledgements}
This research has made use of data obtained from the {\sl Suzaku} satellite, a collaborative mission between the space agencies of Japan (JAXA) and the USA (NASA). We acknowledge the use of public data from the {\sl Swift} data archive.

\appendix
\section{}
Detailed tables of spectra parameters associated with fits performed in Section 3.2 and Section 3.4:
\begin{table*}
\caption{Initial parametrisation of the 0.6-100.0\,keV spectrum, modelled using powerlaw, warm absorption components, \textsc{compTT}, \textsc{pexrav} and Gaussians (all where appropriate). $^{a}$ \textsc{powerlaw} normalization given in units $(10^{-2}\,{\rm ph\,keV^{-1}\,cm^{-2}\,s^{-1}})$. $^{b}$ Flux for \textsc{compTT} quoted over the 0.6-10.0\,keV range in units 10$^{-11}$erg\,cm$^{-2}$\,s$^{-1}$. $^{c}$ Flux given in units $(10^{-5}\,{\rm ph\,cm^{-2}\,s^{-1}})$. $^{d}$ Column density measured in units $10^{22}$\,${\rm cm}^{-2}$. $^{e}$ Ionization parameter given in units erg\,cm\,s$^{-1}$. * Denotes a frozen parameter. $\dagger$ Emission line has been confined to the range 6.63-6.70\,keV.}
\begin{tabular}{l l c c c c c}
\hline
Feature & Parameter & Fairall 9 & MCG--6-30-15 & NGC 3516 & NGC 3783 & NGC 4051 \\
\hline
\multirow{2}{*}{Powerlaw} & $\Gamma$ & $2.06^{+0.03}_{-0.03}$ & $2.04^{+0.02}_{-0.02}$ & $1.73^{+0.03}_{-0.06}$ & $1.87^{+0.05}_{-0.01}$ & $2.09^{+0.03}_{-0.03}$ \\ [0.8ex]
& Norm$^{a}$ & $0.92^{+0.02}_{-0.03}$ & $1.73^{+0.04}_{-0.04}$ & $0.34^{+0.02}_{-0.02}$ & $0.81^{+0.01}_{-0.02}$ & $1.05^{+0.03}_{-0.02}$ \\
\hline
\multirow{4}{*}{\textsc{compTT}} & kT (keV) & $8.9^{+2.3}_{-6.2}$ & $<15.9$ & -- & $8.3^{+0.2}_{-0.1}$ & $3.7^{+1.9}_{-0.8}$ \\ [0.8ex]
& $\tau$ & $0.4^{+0.2}_{-0.2}$ & $<0.2$ & -- & $1.2^{+0.1}_{-0.1}$ & $0.3^{+0.7}_{-0.2}$ \\ [0.8ex]
& Flux\,$^{b}$ & $0.29^{+0.01}_{-0.01}$ & $0.27^{+0.03}_{-0.03}$ & -- & $1.86^{+0.15}_{-0.28}$ & $0.40^{+0.02}_{-0.02}$ \\ [0.8ex]
& $\triangle\chi^{2}$ & -295 & -30 & -- & -33 & -707 \\
\hline
Reflection & $R$ & $1.55^{+0.26}_{-0.24}$ & $0.79^{+0.12}_{-0.08}$ &  $0.87^{+0.32}_{-0.22}$ & $1.04^{+0.04}_{-0.04}$ & $1.52^{+0.26}_{-0.29}$ \\
\hline
\multirow{4}{*}{Fe\,K$\alpha$ core} & LineE (keV) & $6.39^{+0.01}_{-0.01}$ & $6.39^{+0.01}_{-0.01}$ & $6.40^{+0.01}_{-0.01}$ & $6.39^{+0.01}_{-0.01}$ & $6.40^{+0.01}_{-0.01}$ \\ [0.8ex]
& $EW$ (eV) & $75^{+8}_{-8}$ & $30^{+3}_{-3}$ & $229^{+10}_{-52}$ & $101^{+34}_{-10}$ & $67^{+7}_{-7}$ \\ [0.8ex]
& Flux$^{c}$ & $2.21^{+0.23}_{-0.24}$ & $1.39^{+0.16}_{-0.13}$ & $4.16^{+0.19}_{-0.94}$ & $7.93^{+0.27}_{-0.78}$ & $1.71^{+0.18}_{-0.18}$  \\ [0.8ex]
& $\triangle\chi^{2}$ & -337 & -331 & -1396 & -1366 & -320 \\
\hline
Fe\,K$\beta$ & Norm$^{c}$ & 0.29* & 0.18* & 0.53* & 1.03* & 0.22*\\
\hline
\multirow{4}{*}{Fe\,{\rm XXV} emission} & LineE (keV) & $6.66^{+0.04}_{-0.05}$ & $6.63^{+0.04}
$\,$\dagger$ & $6.60^{+0.05}_{-0.05}$ & -- & $6.65^{+0.08}_{-0.04}$ \\ [0.8ex]
& $EW$ (eV) & $15^{+7}_{-7}$ & $12^{+4}_{-3}$ & $19^{+8}_{-8}$ & -- & $33^{+9}_{-8}$ \\ [0.8ex]
& Flux\,$^{c}$ & $0.44^{+0.21}_{-0.22}$ & $0.52^{+0.19}_{-0.12}$ & $0.42^{+0.18}_{-0.17}$ & -- & $0.93^{+0.27}_{-0.22}$ \\ [0.8ex]
& $\triangle\chi^{2}$ & -10 & -51 & -16 & -- & -57 \\
\hline
\multirow{4}{*}{Fe\,{\rm XXVI} emission} & LineE (keV) & -- & 6.97* & -- & 6.97* & -- \\ [0.8ex]
& $EW$ (eV) & -- & $15^{+5}_{-3}$ & -- & $13^{+1}_{-1}$ & --\\ [0.8ex]
& Flux\,$^{c}$ & -- & $0.56^{+0.018}_{-0.11}$ & -- & $8.50^{+0.31}_{-0.31}$ & -- \\ [0.8ex]
& $\triangle\chi^{2}$ & -- & -6 & -- & -17 & -- \\
\hline
\multirow{5}{*}{Broad line} & LineE (keV) & $6.36^{+0.33}_{-0.40}$ & $5.93^{+0.07}_{-0.14}$ & $6.10^{+0.26}_{-0.33}$ & $6.04^{+0.13}_{-0.05}$ & 6.4* \\ [0.8ex]
& $\sigma_{\rm Broad}$ (keV) & $0.81^{+0.43}_{-0.18}$ & $0.84^{+0.06}_{-0.06}$ & $<0.88$ & $0.21^{+0.12}_{-0.05}$ & $0.96^{+0.76}_{-0.38}$ \\ [0.8ex]
& $EW_{\rm Broad}$ (eV) & $62^{+35}_{-30}$ & $149^{+21}_{-9}$ & $55^{+29}_{-28}$ & $30^{+9}_{-6}$ & $87^{+67}_{-46}$ \\ [0.8ex]
& Flux$^{c}$ & $2.00^{+1.12}_{-0.96}$ & $1.31^{+0.68}_{-0.66}$ & $2.42^{+0.74}_{-0.46}$ & $2.43^{+1.87}_{-1.29}$ & $2.43^{+1.87}_{-1.29}$ \\ [0.8ex]
& $\triangle\chi^{2}$ & -13 & -302 & -17 & -55 & -10 \\
\hline 
\multirow{4}{*}{Fe\,{\rm XXV} absorption} & LineE (keV) & -- & $6.76^{+0.02}_{-0.02}$ & -- & $6.72^{+0.03}_{-0.03}$ & $6.79^{+0.07}_{-0.05}$ \\ [0.8ex]
& $EW$ (eV) & -- & $-(27^{+3}_{-3})$ & -- & $-(14^{+5}_{-4})$ & $-(15^{+5}_{-5})$  \\ [0.8ex]
& Flux\,$^{c}$ & -- & $-(1.12^{+0.12}_{-0.11})$ & -- & $-(1.02^{+0.28}_{-0.39})$ & $-(0.47^{+0.17}_{-0.17})$ \\ [0.8ex]
& $\triangle\chi^{2}$ & -- & -62 & -- & -33 & -11 \\
\hline
\multirow{4}{*}{Fe\,{\rm XXVI} absorption} & LineE (keV) & -- & $7.05^{+0.02}_{-0.02}$ & -- & -- & $7.12^{+0.03}_{-0.03}$ \\ [0.8ex]
& $EW$ (eV) & -- & $-(21^{+3}_{-4})$ & -- & -- & $-(27^{+7}_{-7}$) \\ [0.8ex]
& Flux\,$^{c}$ & -- & $-(0.75^{+0.13}_{-0.11})$ & -- & -- & $-(0.55^{+0.14}_{-0.14})$ \\ [0.8ex]
& $\triangle\chi^{2}$ & -- & -81 & -- & -- & -47 \\
\hline
\multirow{2}{*}{Warm absorption zone 1} & $N_{\rm H}$\,$^{d}$ & -- & $0.23^{+0.04}_{-0.02}$ & $3.21^{+0.55}_{-1.65}$ & $0.07^{+0.14}_{-0.06}$ & $0.57^{+0.47}_{-0.18}$ \\ [0.8ex]
& log$(\xi)$\,$^{e}$ & -- & $0.79^{+0.06}_{-0.09}$ & $2.15^{+0.03}_{-0.05}$ & $2.10^{+0.04}_{-0.01}$ & $3.01^{+0.22}_{-0.03}$ \\
\hline
\multirow{2}{*}{Warm absorption zone 2} & $N_{\rm H}$\,$^{d}$ & -- & $0.70^{+0.09}_{-0.06}$ & -- & $0.32^{+0.04}_{-0.05}$ & $0.09^{+0.07}_{-0.04}$ \\ [0.8ex]
& log$(\xi)$\,$^{e}$ & -- & $1.84^{+0.05}_{-0.05}$ & -- & $0.68^{+0.12}_{-0.57}$ & $1.84^{+0.06}_{-009}$ \\
\hline
\textsc{wabs} & $N_{\rm H}$\,$^{d}$ & 0.0316* & 0.0392* & $0.158^{+0.038}_{-0.018}$ & 0.0991* & 0.0115* \\
\hline
XIS/BAT & Const & $0.75^{+0.06}_{-0.05}$ & $0.87^{+0.05}_{-0.04}$ & $1.74^{+0.08}_{-0.07}$ & $0.82^{+0.02}_{-0.02}$ & $0.69^{+0.05}_{-0.05}$ \\
\hline
& $\chi^{2}_{\nu}$ & 958.7/897 & 2085.3/1820 & 537.7/497 & 1528.6/1376 & 1136.0/1085  \\
\hline
\end{tabular}
\label{tab:parametrisation}
\end{table*}

\begin{table*}
\caption{Components for the baseline model (without broad or diskline emission) to the long observations with {\sl Suzaku} XIS, HXD and BAT data from {\sl Swift}. $^{a}$ \textsc{powerlaw} normalization given in units $(10^{-2}\,{\rm ph\,keV^{-1}\,cm^{-2}\,s^{-1}})$. $^{b}$ Column density for partial coverer measured in units $10^{23}$\,${\rm cm}^{-2}$. $^{c}$ Ionization parameter given in units erg\,cm\,s$^{-1}$. $^{d}$ Flux for \textsc{compTT} quoted over the 0.6-10.0\,keV range and \textsc{reflionx} over the 2.0-100.0\,keV range in units 10$^{-11}$erg\,cm$^{-2}$\,s$^{-1}$. $^{e}$ \textsc{reflionx} normalisation given in units $10^{-5}$. $^{f}$ Flux given in units $(10^{-5}\,{\rm ph\,cm^{-2}\,s^{-1}})$. $^{g}$ Column density measured in units $10^{22}$\,${\rm cm}^{-2}$. * indicates a frozen parameter. The highly ionized zone for NGC 3783 has redshift fixed at $z=0.0097$.}
\begin{tabular}{l l c c c c c}
\hline
Component & Parameter & Fairall 9 & MCG--6-30-15 & NGC 3516 & NGC 3783 & NGC 4051 \\
\hline
\multirow{2}{*}{Powerlaw} & $\Gamma$ & $1.94^{+0.02}_{-0.03}$ & $1.87^{+0.04}_{-0.04}$ & $1.75^{+0.04}_{-0.06}$ & $1.79^{+0.02}_{-0.13}$ & $2.04^{+0.04}_{-0.05}$ \\ [0.8ex]
& Norm\,$^{a}$ & $0.88^{+0.03}_{-0.03}$ & $1.62^{+0.05}_{-0.04}$ & $0.34^{+0.02}_{-0.03}$ & $1.52^{+0.02}_{-0.11}$ & $0.95^{0.04}_{-0.05}$ \\ 
\hline
Absorbed powerlaw & Norm\,$^{a}$ & -- & $1.63^{+0.16}_{-0.33}$ & $0.06^{+0.04}_{-0.04}$ & -- & $0.55^{+0.14}_{-0.15}$ \\ [0.8ex]
\multirow{3}{*}{Partial coverer} & $N_{\rm H}$\,$^{b}$ & -- & $34.07^{+2.38}_{-3.77}$ & $1.15^{+3.02}_{-0.75}$ & -- & $35.91^{+5.92}_{-6.15}$ \\ [0.8ex]
& log$(\xi)$\,$^{c}$ & -- & $2.4^{+0.1}_{-0.1}$ & $3.1^{+0.5}_{-0.5}$ & -- & $2.6^{+0.3}_{-0.6}$\\ [0.8ex]
& $C_{\rm frac}$ & -- & $\sim50\%$ & $\sim21\%$ & -- & $\sim37\%$ \\
\hline
\multirow{4}{*}{\textsc{compTT}} & kT (keV) & $<4.1$ & $<8.8$ & -- & $8.5^{+0.2}_{-0.1}$ & $5.4^{+3.1}_{-1.0}$ \\ [0.8ex]
& $\tau$ & $0.8^{+1.4}_{-0.6}$ & $0.8^{+12.9}_{-0.1}$ & -- & $1.6^{+0.1}_{-0.1}$ & $0.9^{+0.4}_{-0.2}$ \\ [0.8ex]
& Flux\,$^{d}$ & $0.50^{+0.14}_{-0.14}$ & $0.72^{+0.18}_{-0.10}$ & -- & $2.24^{+5.31}_{-0.76}$ & $0.49^{+0.08}_{-0.08}$ \\ [0.8ex]
& $\triangle\chi^{2}$ & -413 & -74 & -- & -461 & -333 \\
\hline
\multirow{4}{*}{\textsc{reflionx}} & $\xi^{c}$ & $<11$ & $<12$ & $<12$ & $<14$ & $<21$ \\ [0.8ex]
& $Z_{\rm Fe}$ & $2.0^{+1.5}_{-0.3}$ & 1.0* & 1.0* & $0.8^{+0.1}_{-0.1}$ & $0.5^{+0.4}_{-0.4}$ \\ [0.8ex]
& Norm$^{e}$ & $2.53^{+0.34}_{-0.33}$ & $1.02^{+0.09}_{-0.28}$ & $1.47^{+0.23}_{-0.43}$ & $5.39^{+1.44}_{-0.26}$ & $0.86^{+0.62}_{-0.42}$\\ [0.8ex]
& Flux$^{d}$ & $2.70^{+0.36}_{-0.35}$ & $1.28^{+0.11}_{-0.35}$ & $2.38^{+0.37}_{-0.70}$ & $8.90^{+2.38}_{-0.43}$ & $1.20^{+0.86}_{-0.58}$ \\
\hline
\multirow{4}{*}{Fe\,{\rm XXV} emission} & LineE (keV) & $6.68^{+0.03}_{-0.03}$ & -- & $6.61^{+0.07}_{-0.07}$ & -- & $6.59^{+0.05}_{-0.07}$\\ [0.8ex]
& $EW$ (eV) & $23^{+6}_{-6}$ & -- & $12^{+8}_{-8}$ & -- & $16^{+7}_{-7}$ \\ [0.8ex]
& Flux\,$^{f}$ & $0.69^{+0.18}_{-0.18}$ & -- & $0.25^{+0.17}_{-0.16}$ & -- & $0.49^{+0.20}_{-0.19}$ \\ [0.8ex]
& $\triangle\chi^{2}$ & -40 & -- & -8 & -- & -21 \\
\hline
\multirow{4}{*}{Fe\,{\rm XXVI} emission} & LineE (keV) & $7.01^{+0.05}_{-0.05}$ & 6.97* & -- & $6.96^{+0.01}_{-0.01}$ & -- \\ [0.8ex]
& $EW$ (eV) & $17^{+8}_{-8}$ & $10^{+4}_{-4}$ & -- & $44^{+11}_{-10}$ & -- \\ [0.8ex]
& Flux\,$^{f}$ & $0.40^{+0.18}_{-0.18}$ & $0.41^{+0.18}_{-0.18}$ & -- & $4.90^{+1.19}_{-1.09}$ & -- \\ [0.8ex]
& $\triangle\chi^{2}$ & -14 & -7 & -- & -45 & -- \\
\hline
\multirow{2}{*}{Warm absorption zone 1} & $N_{\rm H}$\,$^{g}$ & -- & $0.22^{+0.06}_{-0.03}$ & $1.90^{+0.66}_{-0.21}$ & $4.68^{+0.03}_{-0.03}$ & $0.18^{+2.09}_{-0.11}$ \\ [0.8ex]
& log$(\xi)$\,$^{c}$ & -- & $0.76^{+0.13}_{-0.09}$ & $1.98^{+0.13}_{-0.06}$ & $2.10^{+0.01}_{-0.01}$ & $3.01^{+0.37}_{-0.37}$ \\
\hline
\multirow{2}{*}{Warm absorption zone 2} & $N_{\rm H}$\,$^{g}$ & -- & $0.47^{+0.09}_{-0.03}$ & -- & $0.19^{+0.04}_{-0.10}$ & $0.18^{+0.07}_{-0.11}$ \\ [0.8ex]
& log$(\xi)$\,$^{c}$ & -- & $1.76^{+0.05}_{-0.08}$ & -- & $0.69^{+0.11}_{-0.10}$ & $1.88^{+0.07}_{-0.09}$ \\
\hline
\multirow{3}{*}{High $\xi$ zone} & $N_{\rm H}$\,$^{g}$ & -- & $3.99^{+3.65}_{-1.28}$ & -- & $4.30^{+0.80}_{-0.72}$ & $9.55^{+18.95}_{-5.68}$ \\ [0.8ex]
& log$(\xi)$\,$^{c}$ & -- & $3.94^{+0.08}_{-0.25}$ & -- & $3.99^{+0.07}_{-0.08}$ & $4.11^{+0.25}_{-0.15}$ \\ [0.8ex]
& $v_{\rm out}$\,$({\rm km}\,{\rm s}^{-1})$ & -- & $3200^{+400}_{-500}$ & -- & -- & $5600^{+800}_{-700}$ \\
\hline
\textsc{wabs} & $N_{\rm H}$\,$^{g}$ & 0.0316* & 0.0392* & $0.178^{+0.011}_{-0.017}$ & 0.0991* & 0.0115* \\
\hline
XIS/BAT & Const & $0.69^{+0.06}_{-0.06}$ & $0.95^{+0.06}_{-0.07}$ & $1.82^{+0.11}_{-0.08}$ & $0.84^{+0.03}_{-0.02}$ & $0.68^{+0.06}_{-0.05}$ \\
\hline
& $\chi^{2}_{\nu}$ & 959.1/885 & 2026.8/1823 & 521.2/495 & 1493.1/1379 & 1091.5/1086 \\
\hline
\end{tabular}
\label{tab:baseline}
\end{table*}

\begin{table*}
\caption{List of soft X-ray emission and absorption lines included within the baseline model. $^{a}$ Flux given in units $(10^{-5}\,{\rm ph\,cm^{-2}\,s^{-1}})$, negative flux and $EW$ indicate an absorption line.}
\begin{tabular}{l c c c c } 
\hline
LineE (keV) & $EW$\,(eV) & Flux$^{a}$ & $\triangle\chi^{2}$ & Identification \\
\hline
\multicolumn{5}{c}{Fairall 9} \\
$0.66^{+0.01}_{-0.02}$ & $10^{+4}_{-4}$ & $42.07^{+16.77}_{-16.81}$ & -18 & O\,{\rm VIII} Ly$\alpha$ \\
\hline
\multicolumn{5}{c}{MCG--6-30-15} \\
$0.68^{+0.01}_{-0.01}$ & $7^{+1}_{-2}$ & $31.72^{+4.44}_{-9.05}$ & -31 & O\,{\rm VIII} Ly$\alpha$ \\ [0.8ex]
$0.77^{+0.01}_{-0.01}$ & $20^{+5}_{-4}$ & $71.38^{+16.62}_{-13.07}$ & -892 & O\,{\rm VII} RRC \\ [0.8ex]
$0.95^{+0.01}_{-0.01}$ & $4^{+1}_{-1}$ & $5.58^{+1.48}_{-1.31}$ & -70 & Ne\,{\rm IX} resonance \\ [0.8ex]
$2.37^{+0.01}_{-0.01}$ & $-(9^{+1}_{-1})$ & $-(2.65^{+0.33}_{-0.34})$ & -196 & S\,{\rm XV} 1s-2p\\ [0.8ex]
$2.77^{+0.04}_{-0.08}$ & $-(2^{+2}_{-2})$ & $-(0.45^{+0.36}_{-0.35})$ & -9 & S\,{\rm XVI} 1s-2p \\
\hline
\multicolumn{5}{c}{NGC 3516} \\
$0.80^{+0.01}_{-0.01}$ & $17^{+10}_{-9}$ & $11.18^{+6.89}_{-5.84}$ & -7 & O\,{\rm VII} RRC / Fe\,{\rm XVII} 3d-2p\\ [0.8ex]
$0.91^{+0.02}_{-0.23}$ & $9^{+10}_{-8}$ & $3.96^{+4.35}_{-3.52}$ & -13 & Ne\,{\rm IX}\,He$\alpha$ \\ [0.8ex]
$1.44^{+0.02}_{-0.02}$ & $5^{+3}_{-2}$ & $0.78^{+0.42}_{-0.39}$ & -11 & Mg\,{\rm XII}\,Ly$\alpha$ \\ [0.8ex]
$1.56^{+0.02}_{-0.02}$ & $8^{+3}_{-3}$ & $1.03^{+0.37}_{-0.39}$ & -21 & Mg\,{\rm XI}\,He$\beta$ \\
\hline
\multicolumn{5}{c}{NGC 3783} \\ 
$0.63^{+0.01}_{-0.01}$ & $10^{+1}_{-2}$ & $197.10^{+21.19}_{-39.54}$ & -30 & O\,{\rm VIII} Ly$\alpha$ \\ [0.8ex]
$0.70^{+0.01}_{-0.01}$ & $17^{+2}_{-2}$ & $97.03^{+13.55}_{-8.86}$ & -146 & O\,{\rm VII} RRC \\ [0.8ex]
$1.24^{+0.01}_{-0.01}$ & $2^{+1}_{-1}$ & $5.56^{+1.72}_{-2.28}$ & -25 & Ne\,{\rm IX} RRC / Ne\,{\rm X} 1s-3p \\ [0.8ex]
1.47* & $13^{+1}_{-1}$ & $15.65^{+1.59}_{-1.29}$ & -359 & Mg\,{\rm XII} Ly$\alpha$\\ [0.8ex]
$2.39^{+0.01}_{-0.01}$ & $-(13^{+2}_{-2})$ & $-(5.60^{+0.71}_{-0.72})$ & -163 & S\,{\rm XV} 1s-2p\\ [0.8ex]
$3.28^{+0.03}_{-0.02}$ & $6^{+2}_{-2}$ & $1.55^{+0.47}_{-0.52}$ & -26 & Ar\,{\rm XVIII}\,Ly$\alpha$ \\
\hline
\multicolumn{5}{c}{NGC 4051} \\
$2.22^{+0.03}_{-0.03}$ & $4^{+3}_{-2}$ & $0.63^{+0.41}_{-0.38}$ & -16 & Si\,{\rm XII} 1s-3p \\
\hline
\end{tabular}
\label{tab:soft}
\end{table*}

\begin{table*}
\caption{Components for the dual reflector fit to the long observations with {\sl Suzaku} XIS, HXD and BAT data from {\sl Swift}. $^{a}$ \textsc{powerlaw} normalization given in units $(10^{-2}\,{\rm ph\,keV^{-1}\,cm^{-2}\,s^{-1}})$. $^{b}$ Flux for \textsc{compTT} quoted over the 0.6-10.0\,keV range and \textsc{reflionx} over the 2.0-100.0\,keV range in units 10$^{-11}$erg\,cm$^{-2}$\,s$^{-1}$. $^{c}$ Ionization parameter given in units erg\,cm\,s$^{-1}$. $^{d}$ \textsc{reflionx} normalisation given in units $10^{-5}$. $^{e}$ Flux given in units $(10^{-5}\,{\rm ph\,cm^{-2}\,s^{-1}})$. $^{f}$ Column density measured in units $10^{22}$\,${\rm cm}^{-2}$. * Denotes a frozen parameter. $\dagger$ Emission line has been confined to the range 6.63-6.70\,keV. Spin can only be constrained in MCG--6-30-15 and NGC 3783 at the 90\% confidence level. The highly ionized zone for NGC 3783 has redshift fixed at $z=0.0097$ and $v_{\rm out}$ is fixed at the best fit value from the baseline model (Table \ref{tab:baseline}) for MCG--6-30-15 and NGC 4051. The \textsc{compTT} parameters for MCG--6-30-15 have been frozem at their best-fit values from the baseline model.}
\begin{tabular}{l l c c c c c}
\hline
Component & Parameter & Fairall 9 & MCG--6-30-15 & NGC 3516 & NGC 3783 & NGC 4051 \\
\hline
\multirow{2}{*}{Powerlaw} & $\Gamma$ & $2.00^{+0.03}_{-0.01}$ & $2.05^{+0.01}_{-0.01}$ & $1.76^{+0.04}_{-0.03}$ & $1.85^{+0.05}_{-0.01}$ & $2.05^{+0.02}_{-0.04}$ \\ [0.8ex]
& Norm\,$^{a}$ & $0.87^{+0.03}_{-0.03}$ & $1.73^{+0.02}_{-0.02}$ & $0.35^{+0.02}_{-0.01}$ & $1.54^{+0.01}_{-0.04}$ & $0.93^{+0.05}_{-0.05}$ \\ 
\hline
\multirow{4}{*}{\textsc{compTT}} & kT (keV) & $<14.1$ & 3.9* & -- & $>9.5$ & $5.7^{+8.0}_{-2.9}$\\ [0.8ex]
& $\tau$ & $0.5^{+1.6}_{-0.2}$ & 0.8* & -- & $1.9^{+0.1}_{-0.1}$ & $0.2^{+0.4}_{-0.1}$ \\ [0.8ex]
& Flux\,$^{b}$ & $0.36^{+0.02}_{-0.02}$ & 0.72* & -- & $2.59^{+0.17}_{-0.12}$ & $0.49^{+0.04}_{-0.04}$  \\ [0.8ex]
& $\triangle\chi^{2}$ & -449 & -- & -- & -2062 & -251 \\
\hline
\multirow{4}{*}{Unblurred \textsc{reflionx}} & $\xi^{c}$ & $<58$ & $<11$& $<20$ & $<11$ & $<30$ \\ [0.8ex]
& $Z_{\rm Fe}$ & 2.0* & 1.0* & 1.0* & 1.0* & 1.0* \\ [0.8ex]
& Norm$^{d}$ & $0.98^{+1.73}_{-0.54}$ & $0.70^{+0.08}_{-0.10}$ & $0.72^{+0.21}_{-0.43}$ & $4.50^{+0.12}_{-0.33}$ & $0.99^{+0.10}_{-0.44}$\\ [0.8ex]
& Flux$^{b}$ & $2.42^{+4.28}_{-1.34}$ & $0.71^{+0.08}_{-0.10}$ & $1.16^{+0.34}_{-0.69}$ & $6.92^{+0.19}_{-0.51}$ & $1.00^{+0.16}_{-0.65}$ \\
\hline
\multirow{7}{*}{\textsc{kerrconv*reflionx}} & $q$ & $2.7^{+2.5}_{-0.2}$ & $2.3^{+0.2}_{-0.1}$ & $2.4^{+1.0}_{-0.8}$ & $2.8^{+0.4}_{-0.2}$ & $1.9^{+0.5}_{-1.1}$ \\ [0.8ex]
& $a$ & -- & $0.61^{+0.15}_{-0.17}$ & -- & $<0.31$ & --\\ [0.8ex]
& $i^{\circ}$ & $33^{+4}_{-5}$ & $35^{+2}_{-2}$ & $35^{+6}_{-5}$ & $<13$ & $41^{+7}_{-7}$ \\ [0.8ex]
& $\xi^{c}$ & $40^{+29}_{-15}$ & $<11$ & $<24$ & $52^{+8}_{-27}$ & $<18$ \\ [0.8ex]
& $Z_{\rm Fe}$ & 2.0* & 1.0* & 1.0* & 1.0* & 1.0* \\ [0.8ex]
& Norm$^{d}$ & $0.26^{+0.75}_{-0.16}$ & $2.42^{+0.28}_{-0.27}$ & $0.44^{+0.75}_{-0.18}$ & $1.98^{+0.65}_{-0.77}$ & $1.30^{+0.18}_{-0.72}$ \\ [0.8ex]
& Flux$^{b}$ & $0.72^{+2.06}_{-0.44}$ & $2.28^{+0.26}_{-0.25}$ & $1.28^{+2.18}_{-0.52}$ & $2.91^{+0.96}_{-1.13}$ & $1.00^{+0.14}_{-0.76}$ \\
\hline
\multirow{4}{*}{Fe\,{\rm XXV} emission} & LineE (keV) & $6.69^{+0.05}_{-0.05}$ & $6.63^{+0.04}$\,$\dagger$ & -- & -- & -- \\ [0.8ex]
& $EW$ (eV) & $15^{+11}_{-8}$ & $6^{+5}_{-3}$ & -- & -- & -- \\ [0.8ex]
& Flux\,$^{e}$ & $0.43^{+0.30}_{-0.22}$ & $0.25^{+0.20}_{-0.13}$ & -- & -- & -- \\ [0.8ex]
& $\triangle\chi^{2}$ & -14 & -7 & -- & -- & -- \\
\hline
\multirow{4}{*}{Fe\,{\rm XXVI} emission} & LineE (keV) & $7.02^{+0.05}_{-0.04}$ & 6.97* & -- & $6.96^{+0.01}_{-0.01}$ & -- \\ [0.8ex]
& $EW$ (eV) & $20^{+7}_{-7}$ & $45^{+33}_{-19}$ & -- & $44^{+27}_{-9}$ & -- \\ [0.8ex]
& Flux\,$^{e}$ & $0.47^{+0.17}_{-0.17}$ & $2.81^{+2.08}_{-1.18}$ & -- & $3.40^{+2.06}_{-0.73}$ & -- \\ [0.8ex]
& $\triangle\chi^{2}$ & -14 & -10 & -- & -49 & -- \\
\hline
\multirow{2}{*}{Warm absorption zone 1} & $N_{\rm H}$\,$^{f}$ & -- & $0.21^{+0.03}_{-0.03}$ & $2.58^{+0.39}_{-0.24}$ & $4.61^{+0.05}_{-0.02}$ & $0.24^{+0.31}_{-0.15}$ \\ [0.8ex]
& log$(\xi)$\,$^{c}$ & -- & $0.69^{+0.10}_{-0.11}$ & $2.09^{+0.05}_{-0.02}$ & $2.09^{+0.01}_{-0.01}$ & $2.77^{+0.23}_{-0.44}$ \\
\hline
\multirow{2}{*}{Warm absorption zone 2} & $N_{\rm H}$\,$^{f}$ & -- & $0.73^{+0.05}_{-0.06}$ & -- & $0.15^{+0.24}_{-0.08}$ & $0.19^{+0.11}_{-0.14}$ \\ [0.8ex]
& log$(\xi)$\,$^{c}$ & -- & $1.77^{+0.04}_{-0.06}$ & -- & $<0.86$ & $1.89^{+0.06}_{-0.07}$ \\
\hline
\multirow{3}{*}{High $\xi$ zone} & $N_{\rm H}$\,$^{f}$ & -- & $11.01^{+5.10}_{-3.95}$ & -- & $1.59^{+0.56}_{-0.28}$ & $13.21^{+14.52}_{-6.72}$\\ [0.8ex]
& log$(\xi)$\,$^{c}$ & -- & $4.08^{+0.09}_{-0.09}$ & -- & $3.87^{+0.08}_{-0.09}$ & $4.11^{+0.21}_{-0.12}$ \\ [0.8ex]
& $v_{\rm out}$\,$({\rm km}\,{\rm s}^{-1})$ & -- & 3200* & -- & -- & 5600* \\
\hline
\textsc{wabs} & $N_{\rm H}$\,$^{f}$ & 0.0316* & 0.0392* & $0.182^{+0.013}_{-0.013}$ & 0.0991* & 0.0115* \\
\hline
XIS/BAT & Const & $0.71^{+0.06}_{-0.05}$ & $0.84^{+0.06}_{-0.05}$ & $1.84^{+0.10}_{-0.07}$ & $0.84^{+0.02}_{-0.02}$ & $0.77^{+0.05}_{-0.06}$ \\
\hline
& $\chi^{2}_{\nu}$ & 929.3/881 & 2061.4/1823 & 510.9/495 & 1444.8/1378 & 1111.2/1084 \\
\hline
\end{tabular}
\label{tab:dual_reflector}
\end{table*}


\begin{thebibliography}{100}
\bibitem{Anders Grevesse 1989}
Anders E., Grevesse N., Geochimica et Cosmochimica Acta, 1989, 53, 197

\bibitem{Arnaud 1996}
Arnaud K.A., 1996, {\sl Astronomical Data Analysis Software and Systems V}, eds. Jacoby, G., Barnes, J., pg17, ASP Conf. Series Volume 101

\bibitem{Ballantyne Vaughan Fabian 2003}
Ballantyne D.R., Vaughan S., Fabian A.C., 2003, MNRAS, 342, 239

\bibitem{Baumgartner 2010} 
Baumgartner W.H. et al., 2010, ApJS submitted

\bibitem{Berti Volonteri 2008}
Berti E., Volonteri M., 2008, ApJ, 684, 822

\bibitem{Bianchi et al 2004}
Bianchi S., Matt G., Balestra I., Guainazzi M., Perola G.C., 2004, A\&A, 422, 65

\bibitem{Bianchi et al 2010}
Bianchi S., de Angelis I., Matt G., La Parola V., de Rosa A., Grandi P., Jim\'{e}nez Bail\'{o}n E., Piconcelli E., 2010, A\&A, 522, 64

\bibitem{Blandford Znajek 1977}
Blandford R.D., Znajek R.L., 1977, MNRAS, 179, 433

\bibitem{Brenneman Reynolds 2006}
Brenneman L.W., Reynolds C.S., 2006, ApJ, 652, 1028

\bibitem{Brenneman Reynolds 2009}
Brenneman L.W., Reynolds C.S., 2009, ApJ, 702, 1367

\bibitem{Brenneman 2011}
Brenneman L.W., et al., 2011, arXiv1104.1172

\bibitem{Crummy 2006}
Crummy J., Fabian A.C., Gallo L., Ross R.R., 2006, MNRAS, 365, 1067

\bibitem{Dauser 2010}
Dauser T., Wilms J., Reynolds C.S., Brenneman L.W., 2010, MNRAS, 409, 1534

\bibitem{de la Calle Perez}
de la Calle P\'{e}rez I., et al., 2010, A\&A, 524, A50

\bibitem{Dovciak 2004}
Dov\v{c}iak M., Karas V., Yaqoob T., 2004, ApJS, 153, 205 

\bibitem{Emmanoulopoulos 2011}
Emmanoulopoulos D., Papadakis I.E., McHardy I.M., Nicastro F., Bianchi S., Ar\'{e}valo P., 2011, MNRAS, in press

\bibitem{Fabian et al 1989}
Fabian A.C., Rees M.J., Stella L., White N.E., 1989, MNRAS, 238, 729

\bibitem{Fukazawa et al 2009}
Fukazawa Y. 
et al., 2009, PASJ, 61, 17

\bibitem{Gallo 2011}
Gallo L.C., Miniutti G., Miller J.M., Brenneman L.W., Fabian A.C., Guainazzi M., Reynolds C.S., 2011, MNRAS, 411, 607

\bibitem{Garafalo 2009}
Garafalo D., 2009, ApJ, 699, 400

\bibitem{George Fabian 1991}
George I.M., Fabian A.C., 1991, MNRAS, 249, 352

\bibitem{Gierlinski Done 2004}
Gierli\'{n}ski M., Done C., 2004, MNRAS, 349, L7

\bibitem{Hughes Blandford 2003}
Hughes S.A., Blandford R.D., 2003, ApJ, 585, 101

\bibitem{Kalberla et al 2005}
Kalberla P.M.W., Burton W.B., Hartmann D., Arnal E.M., Bajaja E., Morras R., Poppel W.G.L., 2005, A\&A, 440, 775

\bibitem{Kallman 2004}
Kallman T.R., Palmeri P., Bautista M.A., Mendoza C., Krolik J.H., 2004, ApJS, 155, 675

\bibitem{Kaspi 2000}
Kaspi S., Brandt W.N., Netzer H., Sambruna R., Chartas G., Garmire G.P., Nousek J.A., 2000, ApJ, 535, 17

\bibitem{Kataoka 2007}
Kataoka J. et al., 2007, PASJ, 59, 279

\bibitem{King 2005}
King A.E., Lubow A.H., Ogilvie G.I., Pringle J.E., 2005, MNRAS, 363, 49

\bibitem{King Pringle 2007}
King A.E., Pringle J.E., 2007, MNRAS, 377, 25

\bibitem{Koyama et al 2007}
Koyama K. 
et al., 2007, PASJ, 59S, 23

\bibitem{Laor 1991}
Laor A., 1991, ApJ, 376, 90

\bibitem{Liu 2010}
Liu Y., et al., 2010, ApJ, 710, 1228

\bibitem{Lobban 2010}
Lobban A.P., Reeves J.N., Porquet D., Braito V., Markowitz A.G., Miller L., Turner T.J., 2010, MNRAS, 408, 551

\bibitem{Lobban 2011}
Lobban A.P., Reeves J.N, Miller L., Turner T.J., Braito V., Kraemer S.B., Crenshaw D.M., 2011, MNRAS in press

\bibitem{Magdziarz Zdziarski 1995}
Magdziarz P., Zdziarski A.A., 1995, MNRAS, 273, 837

\bibitem{Maiolino et al. 2010}
Maiolino R. et al., 2010, A\&A, 517, 47

\bibitem(Malkan Sargent 1982)
Malkan M.A., Sargent W.L.W., 1982, ApJ, 254, 22

\bibitem{Mangalam 2009}
Mangalam A., Gopal-Krishna, Wiita P.J., 2009, MNRAS, 397, 2216

\bibitem{Markowitz 2007}
Markowitz A.G., et al., 2007, ApJ, 665, 209

\bibitem{Markowitz 2008}
Markowitz A.G. et al., 2008, PASJ, 60S, 277

\bibitem{Markowitz Reeves 2009}
Markowitz A.G., Reeves J.N., 2009, ApJ, 705, 496

\bibitem{McKernan 2007}
McKernan B, Yaqoob T., Reynolds C.S., 2007, MNRAS, 379, 1359

\bibitem{Moderski Sikora Lasota 1998}
Moderski R., Sikora M., Lasota J.-P., 1998, MNRAS, 301, 142

\bibitem{Miniutti 2007}
Miniutti G., et al., 2007, PASJ, 59S, 315

\bibitem{Miniutti 2009}
Miniutti G., Panessa F., De Rosa A., Fabian A.C., Malizia A., Molina M., Miller J.M., Vaughan S., 2009, MNRAS, 398, 255

\bibitem{Miniutti 2010}
Miniutti G., Piconcelli E., Bianchi S., Vignali C., Bozzo E., 2010, MNRAS, 401, 1315

\bibitem{Miller 2008}
Miller L., Turner T.J., Reeves J.N., 2008, A\&A, 483, 437

\bibitem{Miller 2009}
Miller L., Turner T.J., Reeves J.N., 2009, MNRAS, 399, 69

\bibitem{Miller Turner 2009}
Miller L., Turner T.J., 2009, A\&A Review, 17, 47

\bibitem{Morrison  McCammon 1983}
Morrison R., McCammon D., 1983, ApJ, 270, 119

\bibitem{Nardini et al 2010}
Nardini E., Fabian A.C., Reis R.C., Walton D.J., 2010, arXiv:1008.2157v1

\bibitem{Nandra et al 2007}
Nandra K., O'Neill P.M., George I.M., Reeves J.N., 2007, MNRAS, 382, 194

\bibitem{Nardini et al 2010}
Nardini E., Fabian A.C., Reis R.C., Walton D.J., 2011, MNRAS, 410, 1251


\bibitem{Patrick 2011}
Patrick A.R., Reeves J.N., Porquet D., Markowitz A.G., Lobban A.P., Tershima Y., 2011, MNRAS, 411, 2353

\bibitem{Ponti 2009}
Ponti G., et al., 2009, MNRAS, 394, 1487

\bibitem{Porquet et al 2004}
Porquet D., Reeves J.N., O'Brien P., Brinkmann W., 2004, A\&A, 422, 85

\bibitem{Pounds 2004}
Pounds K.A., Reeves J.N., King A.R., Page K.L., 2004, MNRAS, 350, 10

\bibitem{Reeves 2004}
Reeves J.N., Nandra K., George I.M., Pounds K.A., Turner T.J., Yaqoob T., 2004, ApJ, 602, 648

\bibitem{Reeves 2007}
Reeves J.N., et al., 2007, PASJ, 59S, 301

\bibitem{Reynolds 2005}
Reynolds C.S., Brenneman L.W., Garofalo D., 2005, Ap\&SS, 300, 71

\bibitem{Reynolds 2009}
Reynolds C.S., Nowak M.A., Markoff S., Tueller J., Wilms J., Young A.J., 2009, ApJ, 691, 1159

\bibitem{Rezzolla 2008}
Rezzolla L., Barausse E., Dorband E.N., Pollney D., Reisswig C., Seiler J., Husa S., 2008, Phys. Rev. D, 78, 044002

\bibitem{Rivers 2011}
Rivers E., et al., 2011, ApJ, in press

\bibitem{Ross Fabian Ballantyne 2002}
Ross R.R., Fabian A.C., Ballantyne D.R., 2002, MNRAS, 336, 315

\bibitem{Ross Fabian 2005}
Ross R.R., Fabian A.C., 2005, MNRAS, 358, 211

\bibitem{Schmoll 2009}
Schmoll S. 
et al., 2009, ApJ, 703, 2171

\bibitem{Takahashi et al 2007}
Takahashi T. 
et al., 2007, PASJ, 59S, 35

\bibitem{Takahashi et al 2010}
Takahashi T., et al., 2010, SPIE, 7732, 27

\bibitem{Titarchuk 1994}
Titarchuk L., 1994, ApJ, 434, 313

\bibitem{Tombesi 2010}
Tombesi F., Cappi M., Reeves J.N., Palumbo G.G.C., Yaqoob T., Braito V., Dadina M., 2010, A\&A, 521, 57

\bibitem{Turner 2005}
Turner T.J., Kraemer S.B., George I.M., Reeves J.N., Bottorff M.C., 2005, ApJ, 618, 155

\bibitem{Turner 2007}
Turner T.J., Miller L., Reeves J.N., Kraemer S.B., 2007, A\&A, 475, 121

\bibitem{Turner Miller 2009}
Turner T.J., Miller L., 2009, A\&ARv, 17, 47

\bibitem{Turner 2009}
Turner T.J., Miller L., Kraemer S.B., Reeves J.N., Pounds K.A., 2009, ApJ, 698, 99
\bibitem{Volonteri 2005}
Volonteri M. et al., 2005, ApJ, 620, 69

\bibitem{Yaqoob 2005}
Yaqoob T., Reeves J.N., Markowitz A.G., Serlemitsos P.J., Padmanabhan U., 2005, ApJ, 627, 156

\bibitem{Yaqoob 2007}
Yaqoob T., et al., 2007, PASJ, 59S, 283

\bibitem{Young 2005}
Young A.J., Lee J.C., Fabian A.C., Reynolds C.S., Gibson R.R., Canizares C.R., 2005, ApJ, 631, 733

\bibitem{Zycki 2010}
Zycki P.T., Ebisawa K., Niedzwiecki A., Miyakawa T., 2010, PASJ, 62, 1185

\end{thebibliography}
\end{document}